\documentclass[11pt]{article}

\usepackage[final]{acl}
\usepackage{times}
\usepackage{latexsym}
\usepackage[T1]{fontenc}
\usepackage[utf8]{inputenc}
\usepackage{microtype}
\usepackage{inconsolata}
\usepackage{graphicx}
\usepackage{enumitem}
\usepackage{changepage}
\usepackage{booktabs}
\usepackage{tabularx}
\usepackage{pifont}
\usepackage{multirow}
\usepackage{algorithm}
\usepackage{algpseudocode}
\usepackage{amsmath}
\usepackage{amsfonts}
\usepackage{float}
\usepackage{xspace}
\usepackage{hyperref}
\usepackage{cleveref}
\usepackage[table,xcdraw]{xcolor}

\newcommand{\dataname}{AniMINT\xspace} 

\title{Beyond Screenshots: Evaluating VLMs' Understanding of UI Animations}

\author{Chen Liang, Xirui Jiang, Naihao Deng, Eytan Adar, Anhong Guo \\
  University of Michigan \\
  \texttt{\{clumich, xirui, dnaihao, eadar, anhong\}@umich.edu}}

\begin{document}
\maketitle
\begin{abstract}
AI agents operating on user interfaces must understand how interfaces communicate state and feedback to act reliably. As a core communicative modality, animations are increasingly used in modern interfaces, serving critical functional purposes beyond mere aesthetics. Thus, understanding UI animation is essential for comprehensive interface interpretation. However, recent studies of Vision Language Models (VLMs) for UI understanding have focused primarily on static screenshots, leaving it unclear how well these models handle dynamic UI animations.
To address this gap, we created \textbf{\dataname}, a novel dataset of 300 densely annotated UI animation videos.
We systematically evaluate state-of-the-art VLMs on UI animation understanding, including their abilities to perceive the animation effects, identify animation purposes, and interpret animation meaning. Our results show that VLMs can reliably detect primitive motion.
However, their high-level animation interpretation remains inconsistent, with substantial gaps relative to human performance. Finally, we use Motion, Context, and Perceptual Cues (MCPC) to probe factors affecting VLM performance, revealing key bottlenecks and directions for future improvement.
\end{abstract}
\section{Introduction}

Recent work on AI agents has increasingly focused on building systems that can autonomously perceive, reason about, and act within user interfaces (UI) to complete complex tasks on users’ behalf \citep{li-etal-2023-modelscope, deng2023mindweb, zheng2023seeact, liu2024llavaplus, wang2024mobile, zhang2025appagent}. 
In real-world settings, such agents must also develop a rich understanding of user interfaces, including the ways in which interfaces convey system state, provide feedback, and signal available interaction affordances to users.

A central yet underexplored aspect of this understanding is UI animation.  
Animation plays a fundamental role in modern user interface design to convey feedback and information \cite{BayWeiChang1993, BruceThomas2001, BarbaraTversky2002, JeffreyHeer2007, FannyChevalier2016Ani25}. These short yet critical animations serve more than aesthetic or experiential purposes; they are often essential for interpreting both the interface and the user’s interaction with it. For example, the MacOS dock icon bounces for notifications, and the password input box shakes on wrong password.
In many cases, such animations are the primary or only channel through which this information is communicated.
Unlike icons or illustrations, animation is defined more by \emph{movements} that are drawn than by \emph{drawings} that move \cite{RonaldBaecker1990}. 
Because animation’s meaning is typically encoded in motion rather than accompanying graphics, a still image alone is often insufficient to capture its intended message.
Thus, a comprehensive UI understanding must account for both static content and dynamic animations.

\begin{figure*}[t]
  \centering
  \includegraphics[width=\textwidth]{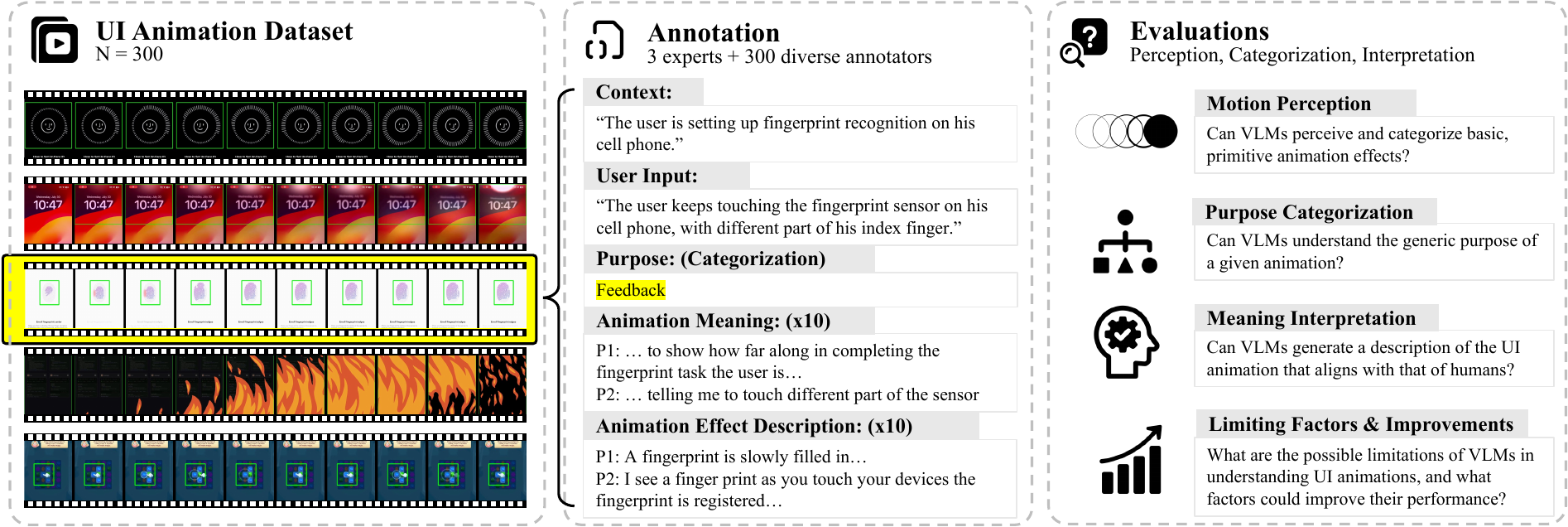}
  \caption{Overview of \dataname, a UI animation dataset with multi-level human annotations. 
  Each clip includes contextual information, an animation purpose label (highlighted in yellow), and ten annotations of the animation’s meaning and effect, supporting evaluations of VLMs across perception, purpose categorization, and interpretation.
}
  \label{fig:main}
\end{figure*}
In this work, we evaluate the capabilities of state-of-the-art VLMs to understand UI animations.
Recent VLMs have shown strong performance on a range of user interface understanding tasks and have been applied to increasingly complex UI-centered problems \cite{PeterShaw2023Pix2Act, JasonWu2024UIClip, OpenAI2025Operator}.
However, to the best of our knowledge, no prior works have systematically studied their capabilities to understand UI animations.

To this end, we constructed \textbf{\dataname}, the \textbf{first} UI \underline{AniM}ation \underline{INT}erpretation dataset.
It contains 300 UI animation videos sourced from web, mobile, and desktop platforms. 
Animations were carefully annotated by 3 UI/UX practitioners and by 300 diverse users, providing a complementary view of UI animations from both experts and general users.
We release our dataset and annotations\footnote{\url{https://github.com/HumanAILab/AniMINT}}.

To systematically evaluate VLMs’ ability to understand UI animations, we formulate a set of research questions based on \dataname that span both low-level animation recognition and higher-level animation understanding.
We evaluated various state-of-the-art VLMs, including models from GPT, Gemini, and other model families.

Our evaluation shows that, although most VLMs reliably recognize primitive animation effects, they struggle with higher-level purpose categorization and meaning interpretation.
To diagnose further, we investigate how Motion, Context, and Perceptual Cues (MCPC) affect UI animation understanding. We augment inputs with motion blending, interaction context, and perceptual captions, then re-evaluate categorization and interpretation. The findings reveal the bottleneck in motion perception, while also highlight the importance of grounding motion in interaction context and higher-level semantic meaning for accurate interpretation.

To summarize, our primary contributions are:
\begin{itemize}[leftmargin=*,itemsep=0pt, topsep=0pt, parsep=0pt]
    \item We introduce \textbf{\dataname}, the first dataset for UI animation understanding, with diverse annotations from both experts and everyday users.
    \item Using \dataname, we conduct a systematic evaluation of nine state-of-the-art VLMs on both primitive animation perception and high-level animation categorization and understanding, revealing substantial gaps in current models' capabilities.
    \item We investigate factors that improve VLMs' capabilities on UI animation understanding and show their effectiveness on Gemini-2.5-Flash.
\end{itemize}

\begin{figure*}[t]
  \centering
  \includegraphics[width=\linewidth]{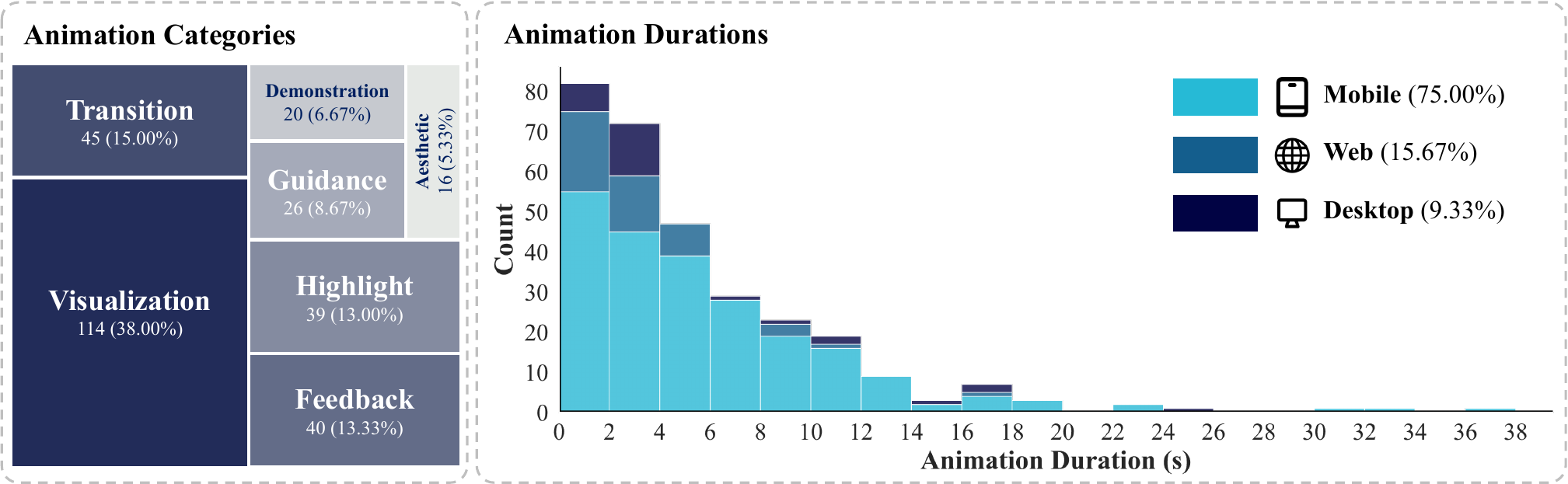}
  \caption{Dataset statistics. 
  (Left) Distribution across seven animation purposes based on prior taxonomy. (Right) Animation duration by platform (mobile, web, and desktop). The median duration is 3.59s.}
  \label{fig:data_stat}
\end{figure*}
\section{Related Work}
\paragraph{UI Animation.} 
Based on \cite{RonaldBaecker1990, Betrancourt2000, FannyChevalier2016Ani25}, we define UI animation as follows to guide data collection for \dataname:

\noindent\textit{UI animation is a deliberately constructed, dynamic transformation of a user interface element that visualizes information or evokes a perceptual or cognitive response in the user.
The transformation extends beyond the immediate next frame.}

UI animations serve functional roles within the interface \cite{BruceThomas2001, BayWeiChang1993, DanielLiddle2016}, such as clarifying state transitions \cite{CharlesEricDessart2011, CelineSchlienger2007}, visualizing information \cite{BarbaraTversky2002, CharlesEricDessart2011, CelineSchlienger2007}, and enhancing user comprehension and experiences \cite{BenediktMerz2016, BruceThomas2001}. 
Drawing from prior literature \cite{RonaldBaecker1990, FannyChevalier2016Ani25, DavidNovick2011, RaquelAvilaMunoz2021Functional}, we categorize animation purposes as Transition, Demonstration, Guidance, Feedback, Visualization, Highlight, and Aesthetic.
We also categorize animation motion effects from the prior literature \cite{BruceThomas2001, DavidNovick2011} into 7 primitive effects, including move, rotate, size, color, fade, blur, and morph. These categorizations guide our data collection and annotation process, and detailed definition can be found in \autoref{ap:rq2_purposes}.

\paragraph{UI Animation Datasets.}
\label{sec_RW_UI_ani_dataset}
There have been datasets that include UI interaction recordings for UI understanding and computer use agent training, such as Rico \cite{Deka2017Rico}, MONDAY \cite{Jang2025Monday}, GUI World \cite{Chen2024Guiworld}, and others across different platforms and tasks \cite{Rawles2023Android, Zhao2025SeeAction, Man2025CAD}. To our knowledge, there is no dataset that specifically focuses on UI animation understanding.
Existing datasets are typically 
designed for specialized tasks such as evaluating state transitions, UI adaptability, or visual signifiers for interaction discoverability \cite{EvaMackamul2025, CharlesEricDessart2011}. Although recordings may include animation, they lack the diversity and annotation needed to evaluate animation understanding. 
In contrast, \dataname~sources diverse UI animation videos annotated by 3 domain experts and 300 general users.

\paragraph{VLMs and VLM Agent in UI Understanding.}
VLMs have emerged as powerful tools across various multimodal tasks, including visual scene comprehension, image captioning, and instructional task execution \cite{grattafiori2024llama, bai2025qwen2}. 
More recently, VLM-based agents have extended these capabilities to interactive settings, enabling models to perceive, reason about, and act within complex visual environments by iteratively grounding language instructions in visual observations \cite{xie2024osworld, wu-etal-2025-webwalker}.
Despite this, their application to UI understanding, particularly regarding dynamic animations, is less explored. 
Existing studies primarily focus on static UI elements, such as visual component identification, interface semantics extraction, and static screen analysis, rather than the dynamic UI properties \cite{RebeccaHenderson2015, MarcusTrapp2013}. 
In this work, we comprehensively evaluate a diverse set of VLMs on UI animation understanding.

\section{\dataname: Dataset for UI \underline{AniM}ation \underline{INT}erpretation}
\paragraph{Dataset and Annotations.}
We crafted a dataset of 300 animation videos collected across mobile, desktop, and web platforms. Mobile animations are mostly collected from the top 100 apps on the App Store and Google Play Store. \autoref{fig:data_stat} visualizes the dataset distribution. The dataset is labeled by 3 domain experts and 300 diverse participants recruited on Prolific. First, each animation is labeled with metadata, including its temporal range, region of interest (ROI), and interaction context. Second, experts assign a purpose category to each animation based on majority voting. Third, we collect open-ended descriptions of each animation’s meaning from general users, obtaining 10 independent responses per animation. In total, this results in 3,000 user-generated descriptions. Participants are compensated \$3 for every 10 responses. The study is IRB approved. Detailed study setup and annotator demographics are provided in \autoref{ap:anno_details}.

\paragraph{Research Questions.}
Based on \dataname, we formulate three research questions to evaluate VLMs' capabilities in understanding UI animations.
Specifically,
Can VLMs perceive and categorize primitive animation effects (\textbf{RQ1}, \Cref{sec_effects})? 
Can VLMs understand the UI animation purpose (\textbf{RQ2}, \Cref{sec_purpose})? 
Can VLMs interpret UI animation meaning (\textbf{RQ3}, \Cref{sec_meaning})? 
Guided by these questions, we further analyze how motion, context, and perceptual cues affect VLM performance to identify key factors for improvement.

\begin{table}[t]
\centering
\small
\begin{tabularx}{\linewidth}{l
    >{\centering\arraybackslash}X
    >{\centering\arraybackslash}X
    >{\centering\arraybackslash}X}
\toprule
\multirow{2}{*}{\textbf{Model Name}} & \multirow{2}{*}{\textbf{Size}} & \textbf{Context Length} & \multirow{2}{*}{\textbf{License}} \\
\midrule
\raisebox{-0.5ex}{\includegraphics[height=1.2em]{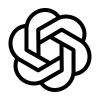}}\,GPT-5               & Unk      & 400K        & Closed \\
\raisebox{-0.5ex}{\includegraphics[height=1.2em]{assets/logos/gpt.png}}\,GPT-5-mini          & Unk     & 400K        & Closed \\
\raisebox{-0.5ex}{\includegraphics[height=1.2em]{assets/logos/gpt.png}}\,GPT-o4-mini         & Unk      & 200K        & Closed \\
\raisebox{-0.5ex}{\includegraphics[height=1.2em]{assets/logos/gpt.png}}\,GPT-o3              & Unk      & 200K        & Closed \\
\raisebox{-0.5ex}{\includegraphics[height=1.2em]{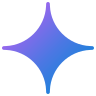}}\,Gemini-2.5-Pro      & Unk      & 1M   & Closed \\
\raisebox{-0.5ex}{\includegraphics[height=1.2em]{assets/logos/gemini.png}}\,Gemini-2.5-Flash    & Unk      & 1M   & Closed \\
\raisebox{-0.5ex}{\includegraphics[height=1.2em]{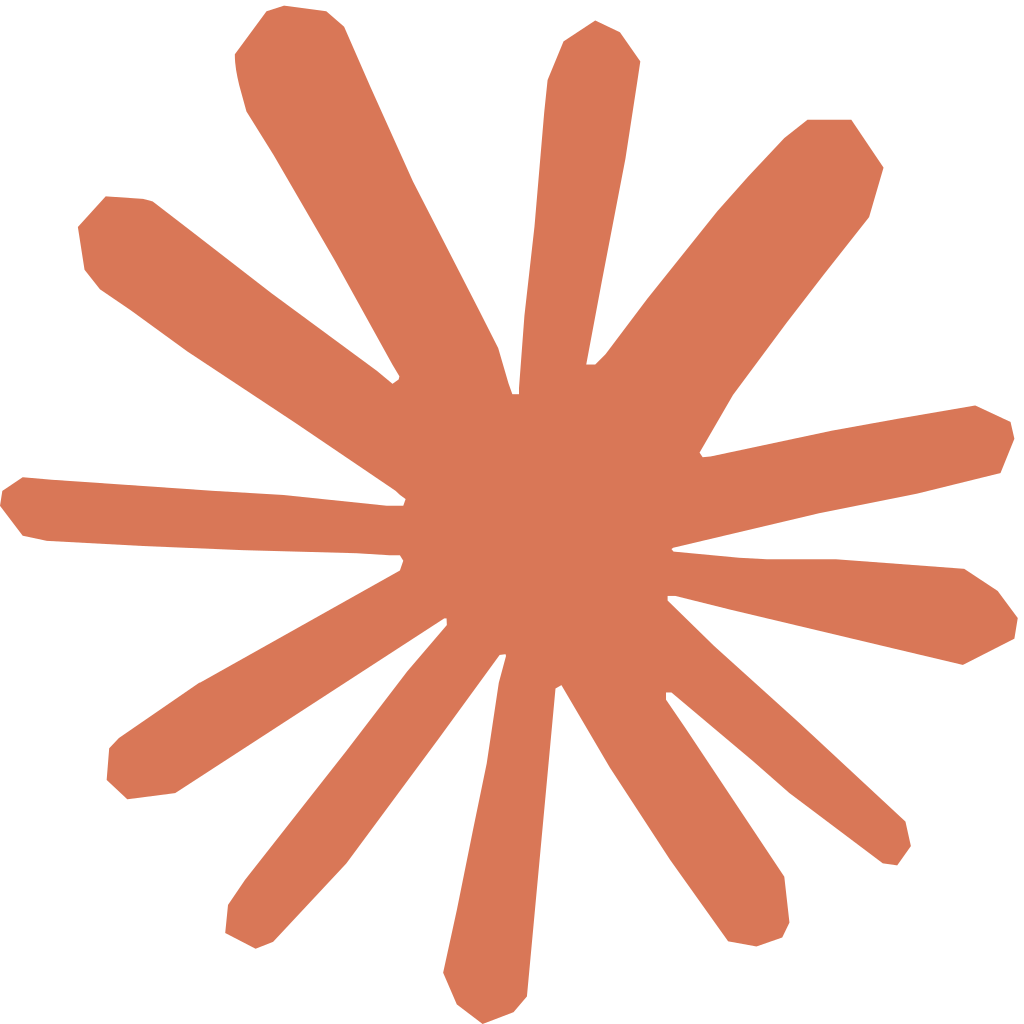}}\,Claude-Sonnet-4     & Unk      & 1M          & Closed \\
\raisebox{-0.5ex}{\includegraphics[height=1.2em]{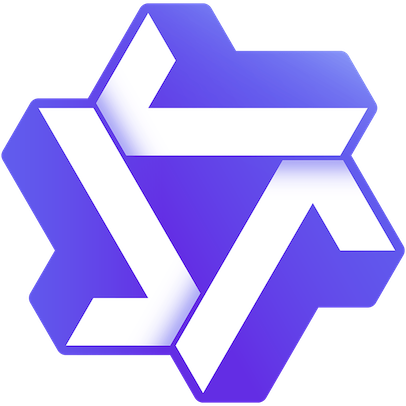}}\,Qwen-2.5-VL-72B     & 73.4B  & 128K   & Open \\
\raisebox{-0.5ex}{\includegraphics[height=1.2em]{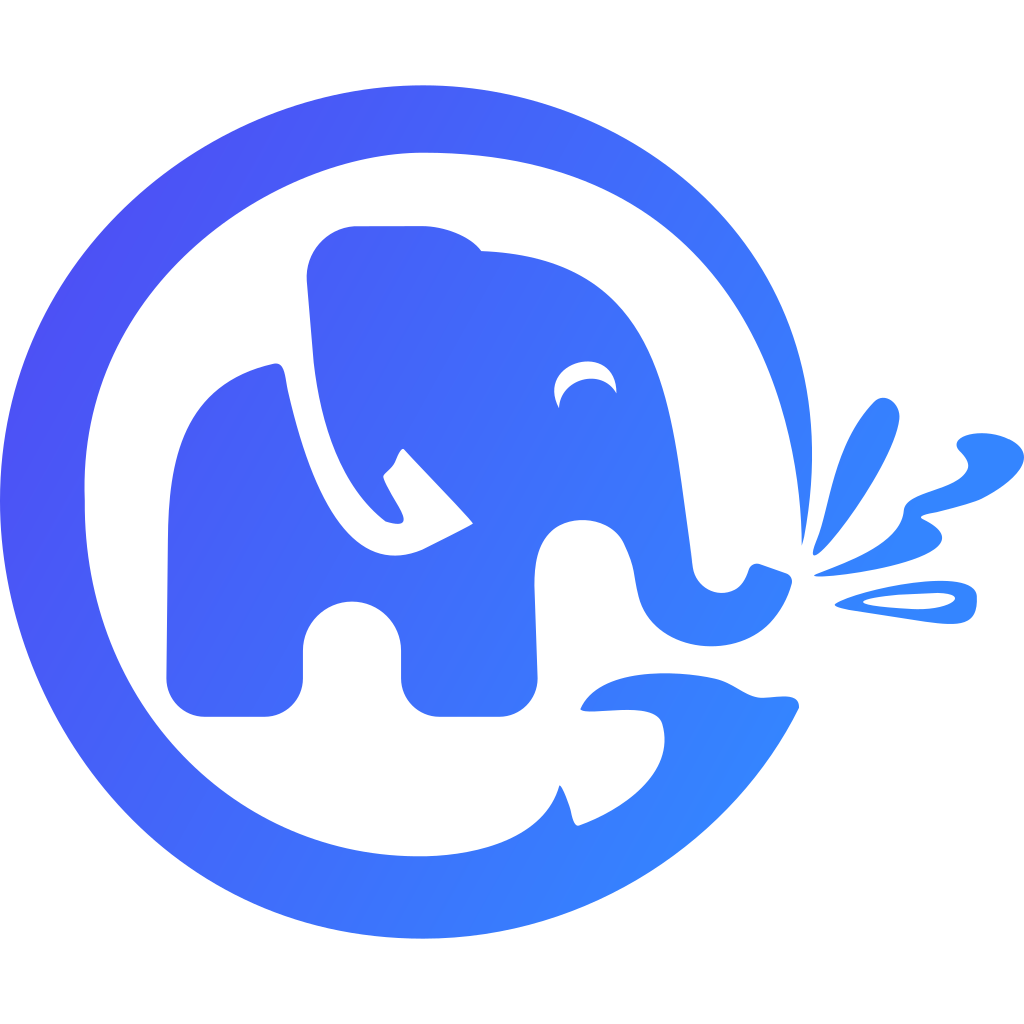}}\,GLM-4.5V            & 106B   & 64K         & Open \\
\bottomrule
\end{tabularx}
\caption{Model details of VLMs tested in this work. }
\label{tab:VLMs}
\end{table}

% \begin{table}[h!]
% \centering
% \begin{tabularx}{\linewidth}{l X X X}
% \toprule
% \textbf{Model Name} & \textbf{Size} & \textbf{Context Length} & \textbf{License Type} \\
% \midrule
% GPT-5               & ? & 400K        & C \\
% GPT-5-mini          & ? & 400K        & C \\
% GPT-o4-mini         & ? & 200K        & C \\
% GPT-o3              & ? & 200K        & C \\
% Gemini-2.5-Pro      & ? & 1,048,576   & C \\
% Gemini-2.5-Flash    & ? & 1,048,576   & C \\
% Claude-Sonnet-4     & ? & 1M          & C \\
% Qwen-2.5-VL-72B     & 73.4B         & 32K--128K   & O \\
% GLM-4.5V            & 106B          & 64k         & O \\
% \bottomrule
% \end{tabularx}
% \caption{Comparison of VLM models by size, context length, and license type.}
% \label{tab:VLMs}
% \end{table}

\paragraph{Model Selections.}
\label{sec:models}
As listed in \Cref{tab:VLMs}, we evaluate 9 state-of-the-art models, including both commercial and open-sourced models. The detailed selection rationale is listed in \autoref{app-sec: model-selection-rationale}. 
We highlight that \dataname~serves as an evaluation similar to \citet{zhou2023instruction, rein2024gpqa}.
Therefore, all experiments are conducted in a zero-shot setting, without any task-specific fine-tuning.

\paragraph{Video Preprocessing.}
We sample the 60 fps source videos at 10 fps, the minimum threshold to avoid significant performance degradation for humans \cite{JessieChen2007}. The sampling strategy is to enable fair comparison among models with and without native video support. More results for video input are in Appendix \ref{app-subsec:rq2-additional-discussion} and \ref{ap:rq3_alignment}. We use green bounding boxes as a visual prompt to highlight the ROIs for models. All frames are resized to a maximum dimension of 480 pixels. This protocol is applied across all models and tasks.

\section{RQ1: Can VLMs Perceive Primitive Animation Effects?}
\label{sec_effects}

\begin{table}[t]
    \small
    \centering
    % Reduce padding to fit column width
    \setlength{\tabcolsep}{3pt} 
    \begin{tabular}{l|c|c|l}
    \toprule
        Motion & \multicolumn{1}{c}{Start} & \multicolumn{1}{c}{End} & Explanation \\
        \midrule
        Blur &  & \includegraphics[width=3.5em]{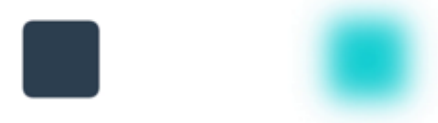} & Changes sharpness or clarity\\
        Color &  &  \includegraphics[width=3.5em]{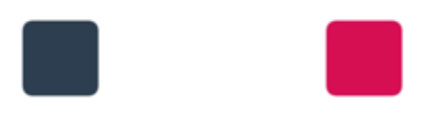} & Modifies hue/saturation/value\\
        Fade &  & \includegraphics[width=3.5em]{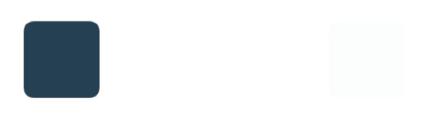} & Adjusts transparency/opacity\\
        Size &  & \includegraphics[width=3.5em]{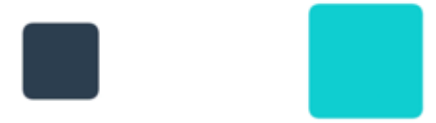} & Scales along an axis\\
        Rotate &  & \includegraphics[width=3.5em]{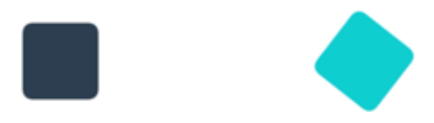} & Rotates orientation\\
        Morph &  & \includegraphics[width=3.5em]{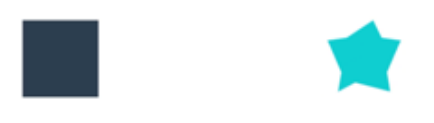} & Transforms shape/form\\
        Move & \multirow{-7}{*}{\includegraphics[width=3.5em]{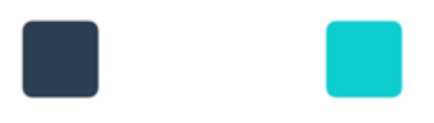}} & \includegraphics[width=3.5em]{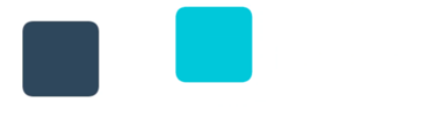} & Changes position\\
        \bottomrule
    \end{tabular}
    \caption{Visual primitives and their effects. A stationary black square (left) serves as a spatial reference.}
    \label{tab:primitive_effects}
\end{table}

\paragraph{Setup.}
We task VLMs with classifying primitive motion sequences into the most representative category among move, rotate, size change, color change, fade, blur, or morph, as shown in \autoref{tab:primitive_effects}. Prompt and video details are listed in \autoref{ap:rq1_primitive}.

\begin{figure}[h]
    \centering
  \includegraphics[width=\linewidth]{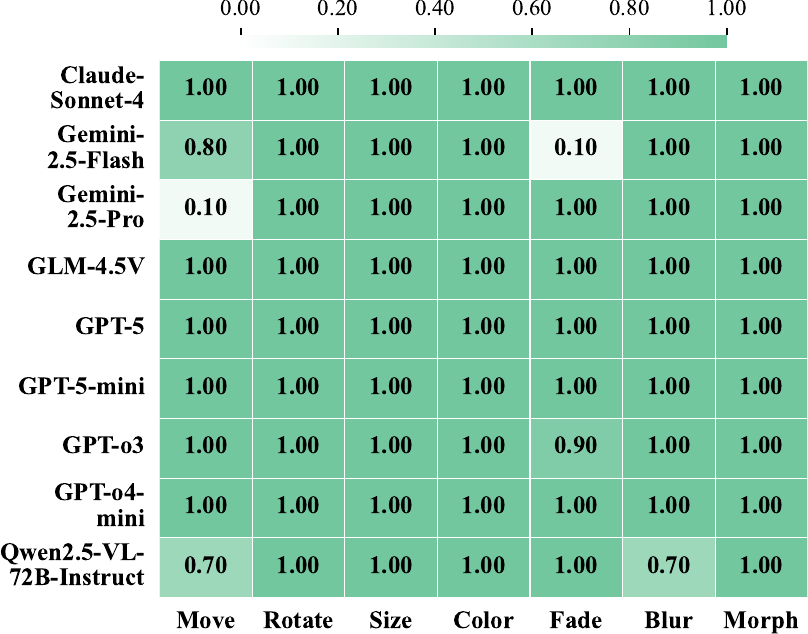}
  \caption{RQ1: VLM accuracy per animation effect.
  To mitigate position bias \citep{zheng2024large}, we average 10 trials per prompt with randomized answer orders, keeping the randomization consistent across all models.
  }
  \label{fig:q1_acc}
\end{figure}

\begin{figure}[t]
    \centering
  \includegraphics[width=\linewidth]{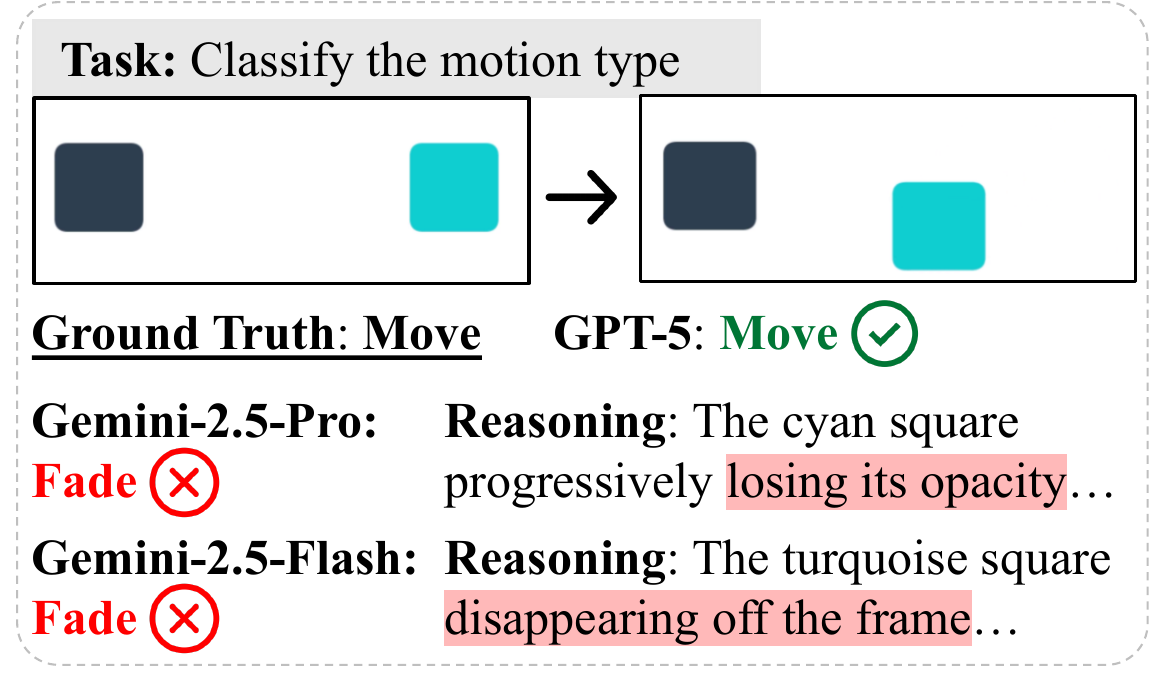}
  \caption{RQ1: An example where the ``move'' motion is incorrectly interpreted as ``fade'' by Gemini-2.5 Pro and Flash.
  These two models hallucinate the motion and reason the object ``progressively losing capacity'' or ``disappearing off the frame.''}
  \label{fig:q1_wrong_cases}
\end{figure}

\paragraph{Answer: Yes.}
Five out of nine models correctly classify all animation effects, including Claude Sonnet 4, GLM-4.5V, GPT-5, GPT-5-mini, and GPT-o4-mini. 
\autoref{fig:q1_acc} reports the corresponding accuracy scores. 
The results indicate that most models capture fundamental motion concepts with only minor errors, such as GPT-o3 misclassifying ``fade'' as ``color change'' in one case.

\subsection{Error Analysis}
\paragraph{Hallucination errors.}
Despite correctly recognizing motion patterns, Gemini-2.5-Pro exhibits hallucination errors. 
In several cases, it describes non-existent visual elements, such as a ``faint, translucent, rounded object'' that does not appear in the animation. 
It also consistently misclassifies ``move'' as ``fade'' and hallucinates that a square is ``progressively losing its opacity'' (\autoref{fig:q1_wrong_cases}). 
These behaviors suggest potential hallucination or misinterpretation of visual cues that aligns with prior observations \citep{li-etal-2023-evaluating, gunjal2024detecting}.

\paragraph{Conceptual confusion.}
Gemini-2.5-Flash shows a pattern where it consistently labels fade as color change, whereas other models selected correctly. 
This suggests difficulty distinguishing subtle differences between closely related animation effects.

\section{RQ2: Can VLMs Understand the UI Animation Purpose?}
\label{sec_purpose}

\paragraph{Setup.}
We task VLMs to categorize the purpose of each animation into one of the seven classes: Transition, Demonstration, Guidance, Feedback, Visualization, Highlight, and Aesthetic. 
\autoref{ap:rq2_purposes} lists the detailed definitions, examples, and the prompt. 
We report classification accuracy and macro-averaged F1 score in \autoref{tab:rq2_model_performance}.

\begin{table}[t]
\centering
\small
\renewcommand*{\arraystretch}{1.3}
\begin{tabular}{lcc}
\toprule
\textbf{Model} & \textbf{Accuracy} & \textbf{\shortstack{Macro F1}} \\
\midrule
\raisebox{-0.5ex}{\includegraphics[height=1.2em]{assets/logos/gemini.png}}\,Gemini-2.5-Pro          & \cellcolor[HTML]{73C79E}\textbf{0.64} & \cellcolor[HTML]{75C89F}\textbf{0.55} \\
\raisebox{-0.5ex}{\includegraphics[height=1.2em]{assets/logos/gpt.png}}\,GPT-5                   & \cellcolor[HTML]{73C79E}0.64          & \cellcolor[HTML]{81CCA8}0.53          \\
\raisebox{-0.5ex}{\includegraphics[height=1.2em]{assets/logos/gpt.png}}\,GPT-o4-mini             & \cellcolor[HTML]{79C9A2}0.63          & \cellcolor[HTML]{8DD1B0}0.51          \\
\raisebox{-0.5ex}{\includegraphics[height=1.2em]{assets/logos/gpt.png}}\,GPT-o3                  & \cellcolor[HTML]{7FCBA6}0.62          & \cellcolor[HTML]{7BCAA4}0.54          \\
\raisebox{-0.5ex}{\includegraphics[height=1.2em]{assets/logos/gemini.png}}\,Gemini-2.5-Flash        & \cellcolor[HTML]{84CEAA}0.61          & \cellcolor[HTML]{81CCA8}0.53          \\
\raisebox{-0.5ex}{\includegraphics[height=1.2em]{assets/logos/gpt.png}}\,GPT-5-mini              & \cellcolor[HTML]{95D4B5}0.58          & \cellcolor[HTML]{9FD9BD}0.48          \\
\raisebox{-0.5ex}{\includegraphics[height=1.2em]{assets/logos/claude.png}}\,Claude-Sonnet-4         & \cellcolor[HTML]{9BD7B9}0.57          & \cellcolor[HTML]{ABDDC5}0.46          \\
\raisebox{-0.5ex}{\includegraphics[height=1.2em]{assets/logos/glm.png}}\,GLM-4.5V                & \cellcolor[HTML]{DEF2E8}0.45          & \cellcolor[HTML]{CFECDE}0.4           \\
\raisebox{-0.5ex}{\includegraphics[height=1.2em]{assets/logos/qwen.png}}\,Qwen2.5-VL-72B-Instruct & \cellcolor[HTML]{FFFFFF}0.39          & \cellcolor[HTML]{FFFFFF}0.32         \\
\bottomrule
\end{tabular}
\caption{RQ2 Model performance comparison. 
Gemini-2.5-Pro achieves the highest accuracy (0.64) in identifying the generic purpose of UI animations, indicating significant room for improvement.
Appendix \ref{app-subsec: rq2-stats-test} provides the pair-wise statistical test. 
}
\label{tab:rq2_model_performance}
\end{table}

\begin{figure}[t]
    \centering
  \includegraphics[width=\linewidth]{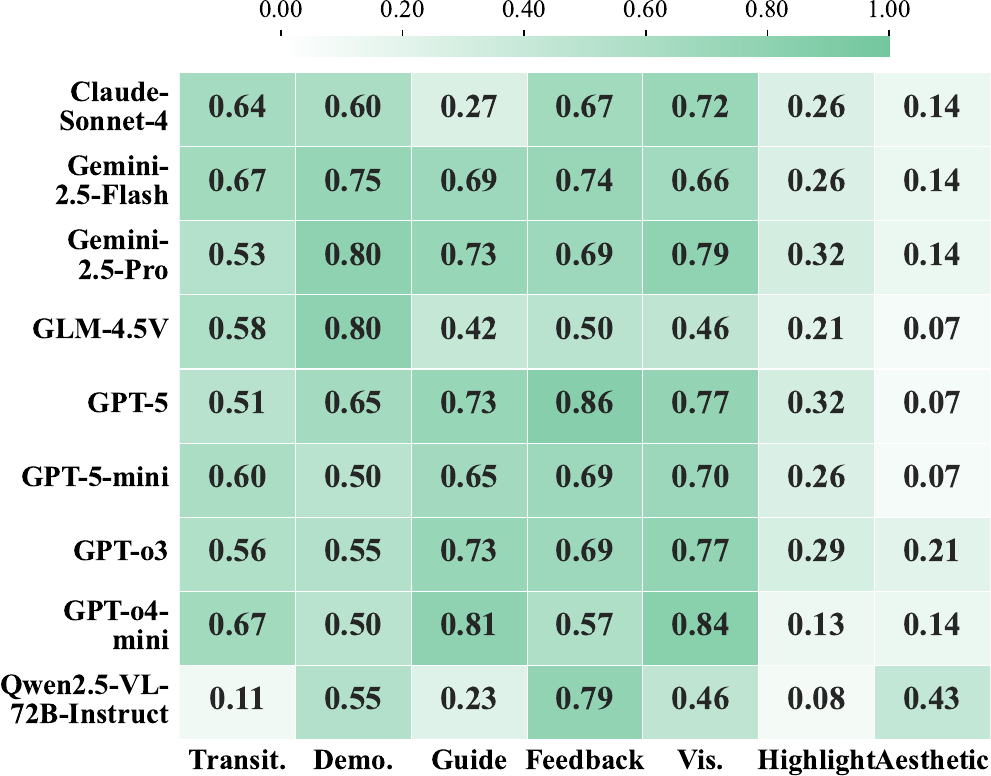}
  \caption{RQ2 per-category recall scores by model and animation purposes.
  While models perform better on animations with more direct functional purposes (such as Transition, Demonstration, Guidance, Feedback, and Visualization), they struggle with animations serving more subtle purposes, such as Highlight and Aesthetic.
  }
  \label{fig:cat_recall}
\end{figure}

\begin{figure*}[t]
    \centering
  \includegraphics[width=\linewidth]{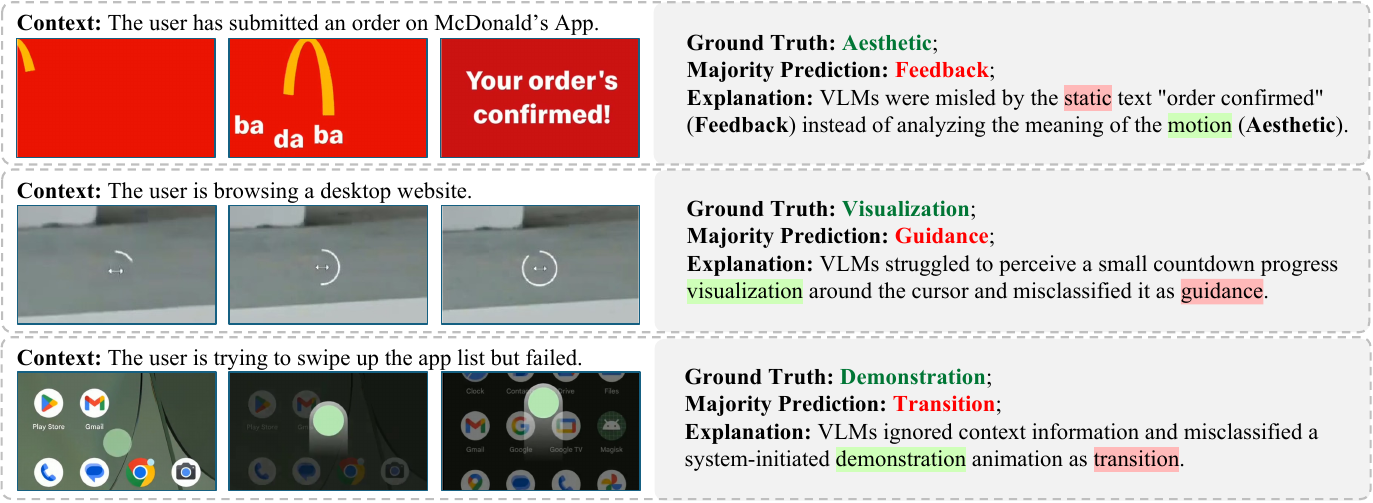}
  \caption{Examples in RQ2 where VLMs fail to identify the correct animation purpose. 
  }
  \label{fig:rq2_failed_cases}
\end{figure*}

\paragraph{Answer: No.}
As shown in \autoref{tab:rq2_model_performance}, the best-performing model, Gemini-2.5-Pro, can only reach an accuracy of 0.64. This shows that VLMs still have a significant gap in understanding the general purpose of UI animations. 

\paragraph{Per-category performance.}
\autoref{fig:cat_recall} shows the per-category recall. Models capture direct functional purposes such as Transition (average recall of 0.54), Demonstration (0.63), Guidance (0.59), Feedback (0.69), and Visualization (0.69) relatively well, with all categories achieving recall above 0.50.
However, performance drops on more subtle purposes, such as Highlight (0.24) and Aesthetic (0.16), which are harder for models to identify.

\paragraph{Per-category difficulty.}
We analyze the majority-vote results across the nine models in \autoref{fig:q2_confusion_matrix}. The models unanimously select the correct label for 56 animations (18.7\%) and reach a correct majority consensus for 176 animations (58.7\%). 
Consistent with the per-category recall results, direct functional categories, particularly Feedback (0.76), Visualization (0.73), and Guidance (0.69), show higher agreement and accuracy than more subtle categories Highlight and Aesthetic.
Additional results and discussions can be found in \autoref{app-subsec:rq2-additional-discussion}.

\subsection{Error Analysis}
\paragraph{VLMs focus on the static frame rather than the animation.}
As shown in \autoref{fig:rq2_failed_cases} (top), six out of nine VLMs incorrectly predict the animation category as Feedback rather than Aesthetic.
The models seem to base their prediction on the final static frame. 
Specifically, the concluding frame displays the message ``Your order is confirmed,'' which conveys feedback to the user. 
However, the animation features the McDonald’s ``M'' logo bouncing into view, accompanied by the text ``ba da ba,'' creating a playful and celebratory effect. 
These motion cues and visual elements serve an aesthetic and emotional purpose, reinforcing brand identity rather than communicating new or necessary information. 
This failure case highlights a limitation of existing VLMs that they tend to overemphasize salient textual cues in static frames rather than interpreting the animation holistically. As a result, visually rich but semantically lightweight animations can be overshadowed by static content, leading to incorrect interpretation.

\paragraph{Small ROIs pose challenges.}
As shown in \autoref{fig:rq2_failed_cases} (middle), VLMs often fail on animations where the countdown progress indicator occupies only a small ROI. Four models ignore the progress bar entirely and instead provide high-level descriptions of the surrounding webpage, and two other models identify an incorrect animation.
Among samples that have a correct majority-voted answer, the mean ROI size is 24.3\% of the screen, which is significantly larger (Mann–Whitney U test, $p = 0.03$) than that of abstained cases (i.e., instances where no consensus among models is reached), whose mean ROI size is 14.1\%.
These errors indicate that when animated elements are visually small, VLMs may fail to localize the relevant motion, leading to incorrect inferences about the animation’s purpose.

\paragraph{VLMs overlook the context.}
The animation shown in \autoref{fig:rq2_failed_cases} (bottom) can, at first glance, be interpreted as a simple transition from the main screen to the app list on Android.
However, the context reveals a different intent: the user repeatedly attempts to swipe up to open the app list but fails to complete the gesture correctly.
In response, the system triggers the animation as a Demonstration to illustrate the correct interaction for the user.
Despite this contextual signal, the VLMs fail to incorporate the context into animation understanding, resulting in eight out of nine models to misclassify the animation as a Transition.
This failure case indicates that current VLMs are not able to well connect perceived UI animations with contextual information such as user intent and prior interaction attempts for interpretation. As a result, animations whose meaning depends on the interaction context are prone to misclassification.

\begin{figure*}[t]
    \centering
  \includegraphics[width=\linewidth]{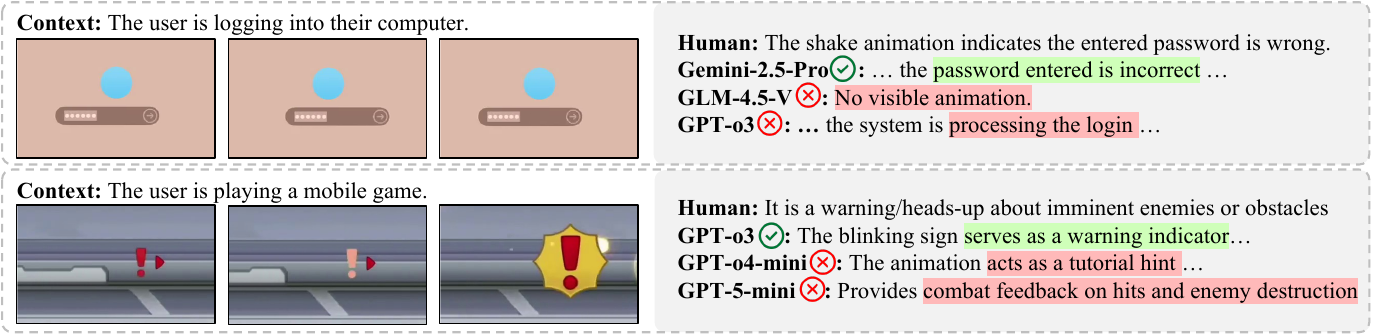}
  \caption{RQ3: Examples of Animation and the interpretations from VLMs.}
  \label{fig:rq3_wrong_example}
\end{figure*}

\section{RQ3: Can VLMs Interpret UI Animations?}
\label{sec_meaning}
\paragraph{Setup.}
We task VLMs to generate a natural language interpretation and compare its semantic similarity to the human responses. 
We use GPT-5-mini as the judge model \cite{ZhengLianmin2023LLMAsJudge}. 
To mitigate the potential bias in LLM-as-a-judge \citep{chen-etal-2024-humans, ye2025justice}, we randomize response orders and prompted the judge model to evaluate independently of length. 
We report the mean and standard deviation of the similarity scores per model. 
For each animation, we leverage the 10 human responses collected and evaluate model predictions either against each individual response or against a summarized version of the responses (details are listed in \autoref{ap:rq3_alignment}).
Since both approaches yield similar model rankings empirically, we report results based on the summarized responses and defer the other results to \autoref{ap:rq3-additional-results}.

\begin{table}[t]
\small
\centering
\renewcommand*{\arraystretch}{1.3}
\resizebox{\linewidth}{!}{
\begin{tabular}{lccc}
\toprule
                        & \multicolumn{1}{l}{Mean ($\uparrow$)} & \multicolumn{1}{l}{Std ($\downarrow$)} & \multicolumn{1}{c}{Distribution} \\
\midrule
\raisebox{-0.5ex}{\includegraphics[height=1.2em]{assets/logos/gpt.png}}\,GPT-o3
  & \cellcolor[HTML]{62C092}{\bf3.47}
  & \cellcolor[HTML]{57BB8A}0.91
  & \raisebox{-0.4ex}{\includegraphics[height=1.5em]{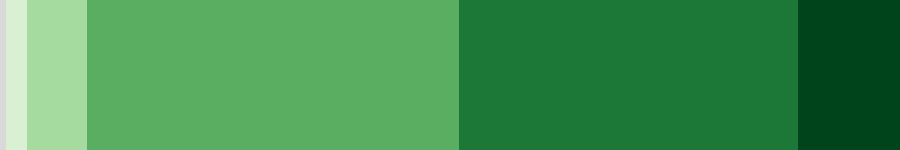}} \\

\raisebox{-0.5ex}{\includegraphics[height=1.2em]{assets/logos/gpt.png}}\,GPT-5
  & \cellcolor[HTML]{68C296}3.44
  & \cellcolor[HTML]{57BB8A}0.90
  & \raisebox{-0.4ex}{\includegraphics[height=1.5em]{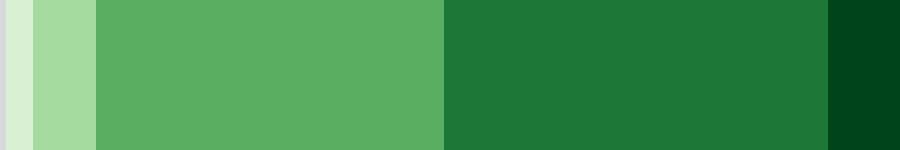}} \\

\raisebox{-0.5ex}{\includegraphics[height=1.2em]{assets/logos/gemini.png}}\,Gemini-2.5-Pro
  & \cellcolor[HTML]{70C69C}3.40
  & \cellcolor[HTML]{57BB8A}0.90
  & \raisebox{-0.4ex}{\includegraphics[height=1.5em]{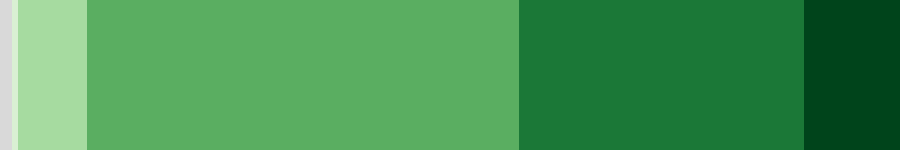}} \\

\raisebox{-0.5ex}{\includegraphics[height=1.2em]{assets/logos/gpt.png}}\,GPT-5-mini
  & \cellcolor[HTML]{70C69C}3.39
  & \cellcolor[HTML]{57BB8A}{\bf0.82}
  & \raisebox{-0.4ex}{\includegraphics[height=1.5em]{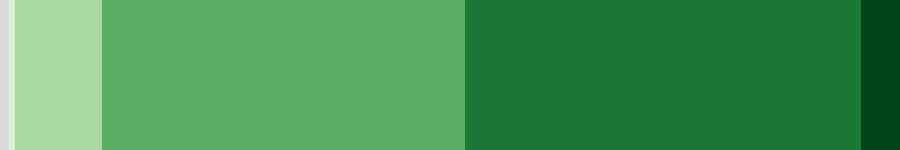}} \\

\raisebox{-0.5ex}{\includegraphics[height=1.2em]{assets/logos/gemini.png}}\,Gemini-2.5-Flash
  & \cellcolor[HTML]{83CDA9}3.31
  & \cellcolor[HTML]{57BB8A}0.95
  & \raisebox{-0.4ex}{\includegraphics[height=1.5em]{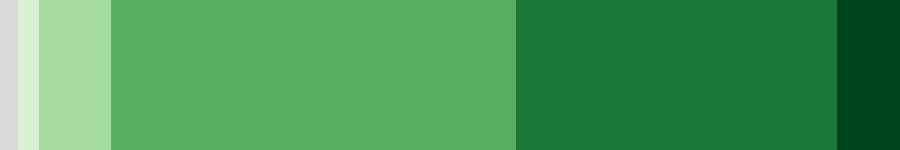}} \\

\raisebox{-0.5ex}{\includegraphics[height=1.2em]{assets/logos/gpt.png}}\,GPT-o4-mini
  & \cellcolor[HTML]{94D4B4}3.23
  & \cellcolor[HTML]{6AC297}1.01
  & \raisebox{-0.4ex}{\includegraphics[height=1.5em]{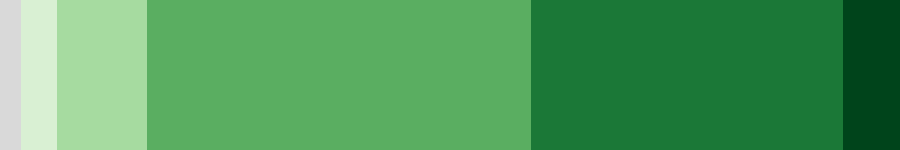}} \\

\raisebox{-0.5ex}{\includegraphics[height=1.2em]{assets/logos/claude.png}}\,Claude-Sonnet-4
  & \cellcolor[HTML]{AFDFC7}3.10
  & \cellcolor[HTML]{8DD1B0}1.12
  & \raisebox{-0.4ex}{\includegraphics[height=1.5em]{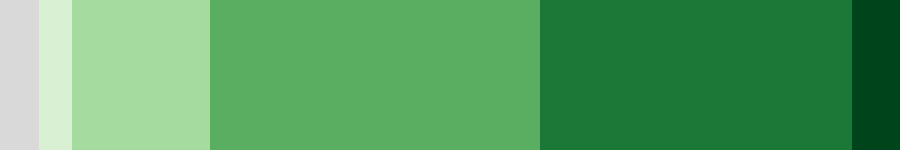}} \\

\raisebox{-0.5ex}{\includegraphics[height=1.2em]{assets/logos/qwen.png}}\,Qwen2.5-VL-72B
  & \cellcolor[HTML]{D0ECDE}2.94
  & \cellcolor[HTML]{B4E0CB}1.24
  & \raisebox{-0.4ex}{\includegraphics[height=1.5em]{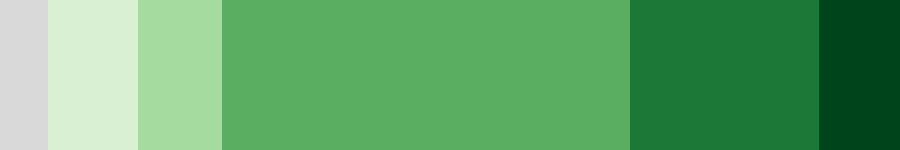}} \\

\raisebox{-0.5ex}{\includegraphics[height=1.2em]{assets/logos/glm.png}}\,GLM-4.5V
  & \cellcolor[HTML]{FFFFFF}2.71
  & \cellcolor[HTML]{FFFFFF}1.47
  & \raisebox{-0.4ex}{\includegraphics[height=1.5em]{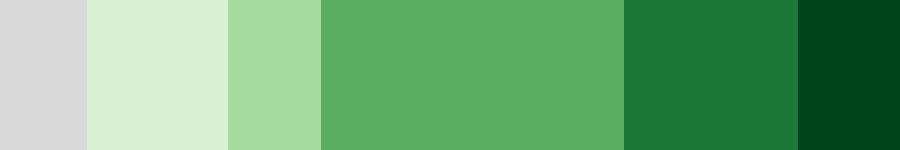}} \\
\bottomrule
\end{tabular}}
\caption{RQ3 Semantic similarity scores between VLM predictions vs. the summarized human response.
We report the score distribution, where the five colors from left to right correspond to scores from 0 to 5.
Appendix \ref{app-subsec: rq3-additional-results} reports results based on individual human responses.
Appendix \ref{app-subsec:rq3-stats-test} provides the pair-wise statistical test. 
}
\label{tab:rq3_vlm_result}
\end{table}

\paragraph{Answer: VLMs capture gist, but miss key details.} 
As shown in \autoref{tab:rq3_vlm_result}, GPT-o3 achieves the highest average score ($3.47\pm0.91$), while GLM-4.5V yields the lowest ($2.71\pm1.47$). 
Most of these VLMs achieve an average score of 3 and above, indicating that VLMs are capable of capturing the gist of animation purposes according to the scoring rubric (Appendix \ref{app-subsec: rq3-eval-prompt}).
However, these VLMs' responses often miss key details or contain subtle differences in nuance.

\subsection{Error Analysis}
\paragraph{Subtle, rapid animations pose challenges.}
For the example in \autoref{fig:rq3_wrong_example} (top), five out of nine models score 0, where they either do not perceive the animation at all (e.g., GLM-4.5v: ``No visible animation''), or hallucinate (e.g., GPT-o3 described a ``collapsing progress bar'' which does not exist).
This animation corresponds to a common UI pattern in which an input box briefly shakes to indicate an incorrect password.
Although this shaking motion is highly recognizable to human users, it is subtle and quick.
As a result, VLMs struggle to detect the motion signal, leading either to missed detections or spurious interpretations.

\paragraph{Small ROIs impact interpretation.}
Similar to RQ2 (\Cref{sec_purpose}), VLMs perform poorly on animations with a small ROI.
As shown in \autoref{fig:rq3_wrong_example} (bottom), a small animated warning indicator, despite exhibiting a visually noticeable animation, receives an average score of 2.
Models that fail on this example either do not perceive the animation at all or are distracted by larger static elements outside the highlighted ROI.
For instance, GPT-5-mini does not mention the red exclamation mark in its reasoning and instead focused on the motion of the jetpack character.
GPT-o4-mini correctly detects the exclamation mark but conflate it with surrounding visual elements (e.g., moving arrows and slider graphics), leading it to misinterpret the animation as a tutorial hint.
These errors suggest that when animated elements are small, VLMs struggle to localize the relevant motion and may default to more visually salient but semantically irrelevant context.
We provide more results on model performance for different categories in 
Appendix \ref{app-subec: rq3-discussion}.

\section{Probing VLM Performance with Motion, Context, and Perceptual Cues}
\label{sec_factors}
\begin{figure}[t]
    \centering
  \includegraphics[width=\linewidth]{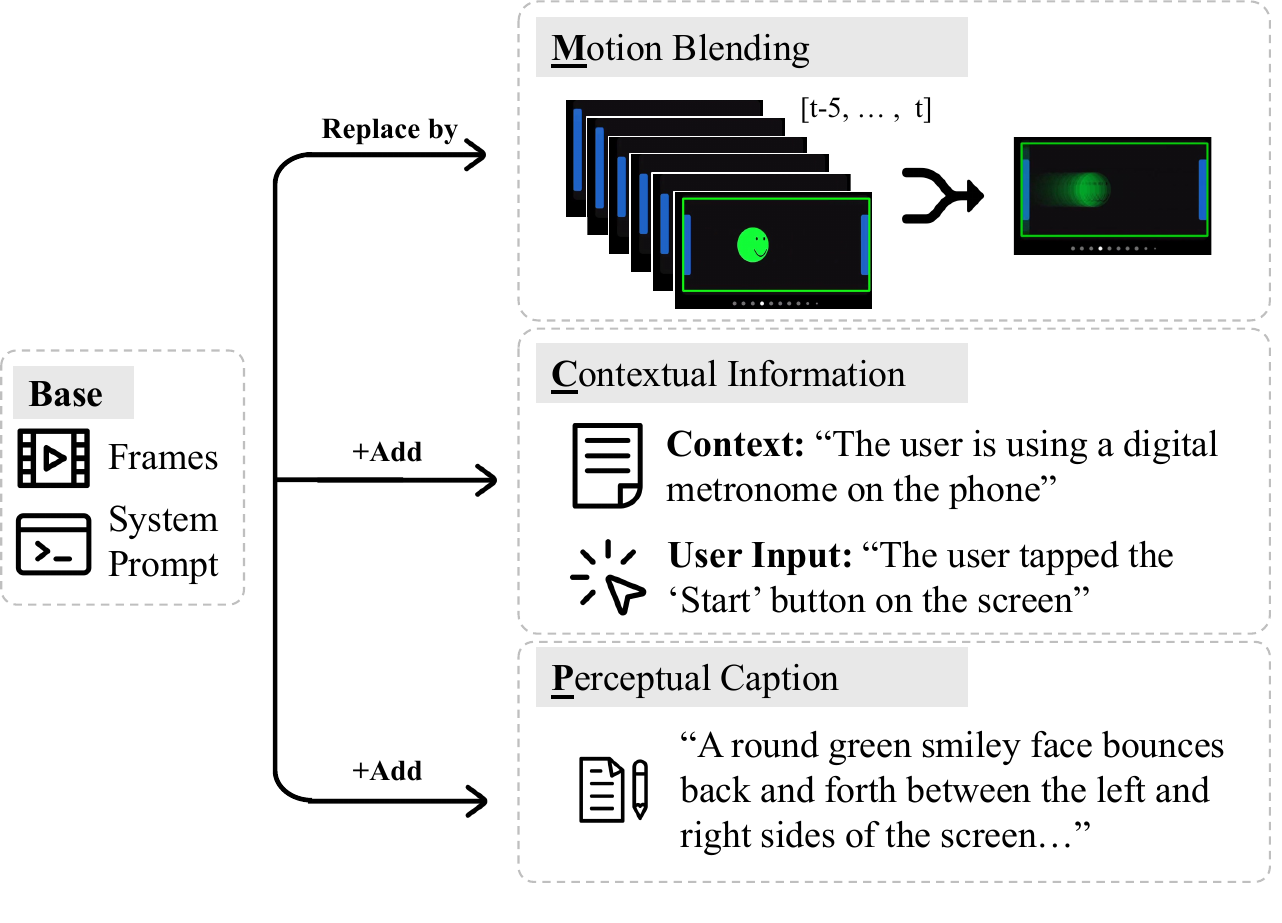}
  \caption{
  MCPC includes Motion Blending (blending the past six frames to capture motion), Contextual Information (interaction context and user input), and Perceptual Caption (textual descriptions of the animation).
  }
  \label{fig:rq4_method}
\end{figure}

\begin{table}[t]
\small
\centering
\renewcommand*{\arraystretch}{1.3}
\resizebox{\linewidth}{!}{
\begin{tabular}{lcccc}
\toprule
                           & \multicolumn{2}{l}{\textbf{Purpose (RQ2)}}                      & \multicolumn{2}{l}{\textbf{Interp (RQ3)}}                \\
                           \midrule
                           & \multicolumn{1}{l}{Acc ($\uparrow$)}               & \multicolumn{1}{l}{F1\textsubscript{Macro} ($\uparrow$)}          & \multicolumn{1}{l}{Mean ($\uparrow$)}              & \multicolumn{1}{l}{Std ($\downarrow$)}               \\
                           \midrule
\raisebox{-0.5ex}{\includegraphics[height=1.2em]{assets/logos/gemini.png}}\,Base      & \cellcolor[HTML]{A1D9BE}0.59          & \cellcolor[HTML]{B0DFC8}0.47          & \cellcolor[HTML]{E8F6EF}3.15          & \cellcolor[HTML]{EAF6F0}1.09          \\
+ M         & \cellcolor[HTML]{EAF7F1}0.52          & \cellcolor[HTML]{E5F5ED}0.41          & \cellcolor[HTML]{FFFFFF}3.08          & \cellcolor[HTML]{E2F3EB}1.07          \\
+ C         & \cellcolor[HTML]{ACDEC5}0.58          & \cellcolor[HTML]{A7DCC2}0.48          & \cellcolor[HTML]{B4E1CB}3.3           & \cellcolor[HTML]{B1DFC8}0.95          \\
+ P         & \cellcolor[HTML]{B6E2CC}0.57          & \cellcolor[HTML]{C2E6D4}0.45          & \cellcolor[HTML]{6FC59B}3.5           & \cellcolor[HTML]{98D5B7}0.89          \\
+ M + P       & \cellcolor[HTML]{E0F3EA}0.53          & \cellcolor[HTML]{EEF8F3}0.4           & \cellcolor[HTML]{76C8A0}3.48          & \cellcolor[HTML]{8CD0AF}0.86          \\
+ M + C       & \cellcolor[HTML]{A1D9BE}0.59          & \cellcolor[HTML]{9ED8BC}0.49          & \cellcolor[HTML]{A6DBC1}3.34          & \cellcolor[HTML]{BDE4D1}0.98          \\
+ C + P       & \cellcolor[HTML]{CBEADB}0.55          & \cellcolor[HTML]{B9E3CE}0.46          & \cellcolor[HTML]{76C8A0}3.48          & \cellcolor[HTML]{67C195}0.77          \\
\bottomrule
{\it +M+C+P} & \cellcolor[HTML]{8CD1AF}\textbf{0.61} & \cellcolor[HTML]{84CDA9}\textbf{0.52} & \cellcolor[HTML]{69C296}\textbf{3.52$^\dagger$} & \cellcolor[HTML]{57BB8A}\textbf{0.73} \\
\bottomrule
\end{tabular}}
\caption{
Effects of augmenting VLM inputs with MCPC on categorization (RQ2) and interpretation (RQ3). We adopt Gemini-2.5-Flash as the backbone model. 
The combined input (last row) outperforms all other combinations, demonstrating the joint effectiveness of MCPC.
Appendix \ref{app-subsec: rq4-stats-test} provides the details of the statistical test. $^\dagger$: significantly better than the base setup.
}
\label{tab:factor_ablation}
\end{table}

\begin{figure}[t]
    \centering
  \includegraphics[width=\linewidth]{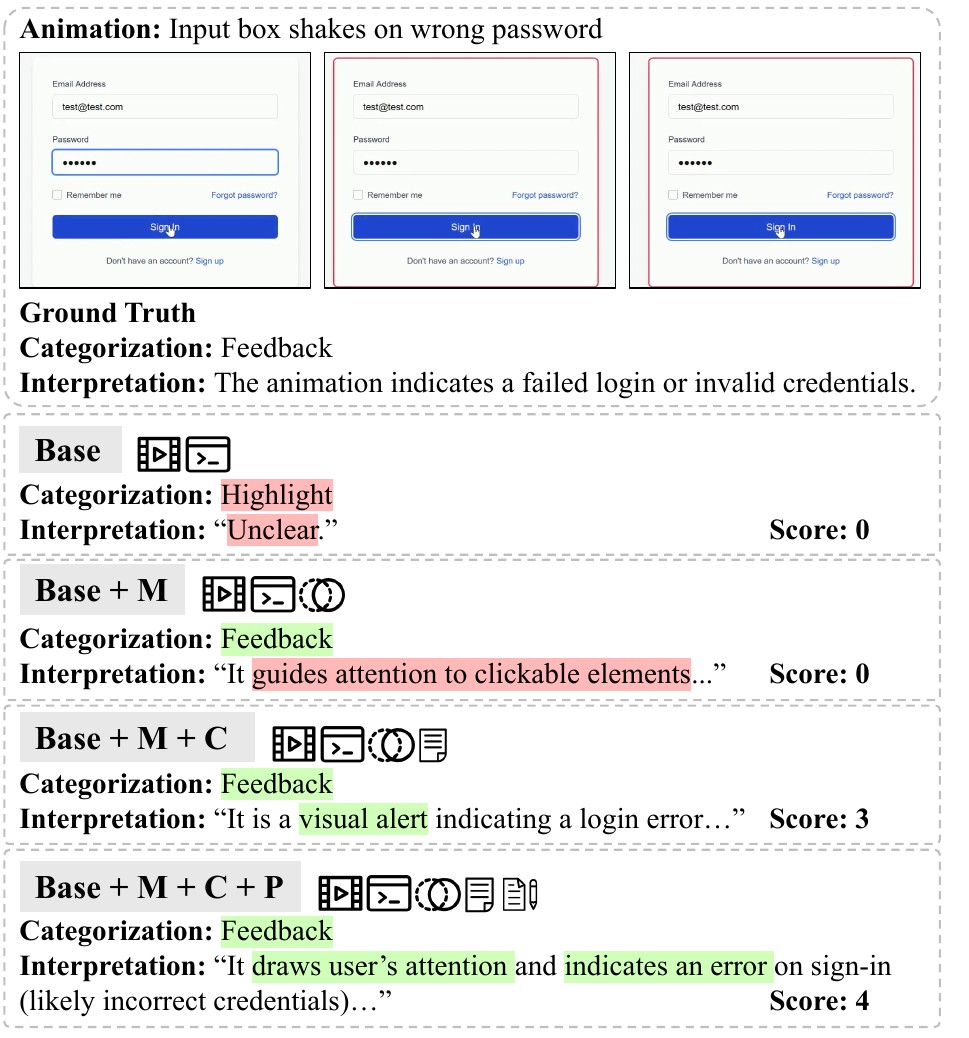}
  \caption{
  Improvements of incorporating MCPC: a wrong password shake is incorrectly classified as highlight in the base condition, but is correctly interpreted as an error indication with MCPC.
  }
  \label{fig:rq4_example}
\end{figure}

\paragraph{Setup.}
To identify limitations and potential improvement factors, we study how Motion, Context, and Perceptual Cues affect VLM performance. (\autoref{fig:rq4_method}). 
For motion blending, we blend past frames into a single image with decaying transparency. This is inspired by Phosphor \cite{PatrickBaudisch2006} that uses afterglow to show UI changes. 
For user context, we add contextual information such as the context of use and users' performed interactions.
For perceptual caption, we provide the annotated text caption of what animations or motions are happening in the video. 
By combining these three factors, we re-evaluate VLM performance on purpose understanding (\Cref{sec_purpose}) and UI animation interpretation (\Cref{sec_meaning}).
We test these combinations using Gemini-2.5-Flash, a lightweight model that demonstrates strong performance in earlier experiments. 
All other experimental setups are kept identical to \Cref{sec_purpose,sec_meaning}. More details about MCPC are listed in \autoref{ap:mcp}.

\paragraph{Results and Discussions.}
As shown in \autoref{tab:factor_ablation}, combining motion, context, and perception cues leads to the best overall performance for both tasks. 
In \autoref{fig:rq4_example}, the vanilla model fails both tasks, misclassifying the animation purpose as ``Highlight'' and describing the meaning as ``unclear.'' 
With motion encoding alone, the model successfully classifies the purpose.
Combining all three factors leads to the best performance, where the model successfully categorizes the shake and provides the most accurate interpretation compared to the other setups.
This demonstrates the importance of motion, context, and perception, as well as the synergy effects across these factors in UI understanding. 

\section{Conclusion}
\label{sec_limitations}
In this paper, we investigate an often overlooked yet critical aspect of UI understanding -- motion and animation. 
We construct \dataname, a densely-annotated UI animation dataset sourced from real-world applications, and comprehensively evaluate a diverse set of state-of-the-art VLMs. 
We find that while most VLMs are capable of perceiving primitive motion effects, they struggle to categorize the animation purpose using the UI animation taxonomy. 
Also, although VLMs' interpretations often capture the gist, they frequently miss key details in their description. 
Furthermore, we investigate performance variations by  
encoding motion cues into images, adding contextual information, and supplying perception captions. They improve VLMs' performance on both the categorization and interpretation tasks, revealing the bottleneck of motion perception and the important synergy effects across perception and semantic context. 
We envision this work and our \dataname{} dataset as a step toward interaction-aware LLM agents that operate between users and interfaces, using UI animation understanding to assist, explain, and guide user interactions involving complex animated behaviors.

\section{Ethical Statement}
This project was conducted in accordance with established ethical standards. All collected data were manually reviewed by the authors to ensure that no sensitive content (e.g., sexual material or violence) or potentially harmful visual stimuli (e.g., rapid flashing) were presented to annotators. Both the video data and the associated annotations were screened to prevent the inclusion of any personally identifiable information. All participants were recruited anonymously, provided informed consent, and were informed of their right to withdraw from the study at any time. The study protocol was approved by the IRB. Additional details regarding the annotation procedure are provided in \autoref{ap:anno_details}.

\section{Limitations}
Due to time and cost constraints, the collected animations are primarily sourced from U.S. based applications where English is the primary language. 
Design practices and interaction patterns may vary across regions due to factors such as language reading direction (e.g., right-to-left vs. left-to-right) and cultural conventions (e.g., shaking to indicate confirmation) \citep{shen-etal-2024-understanding, mogrovejo2024cvqa}.
We acknowledge that including data from a wider range of geographic and cultural contexts could introduce greater diversity into the dataset \citep{mihalcea2025ai}.
However, \dataname{} is the first step in constructing a UI animation understanding dataset.
We encourage future efforts in our community to diversify the animation sources and consider the cultural and language nuances.

Second, in the annotation process, all annotators were recruited within the United States and were English speakers, which may introduce interpretation bias in certain cases. 
Though this happens in many well-known NLP benchmarks \citep{deng2009imagenet, bowman-etal-2015-large} \footnote{These early datasets typically do not report the annotator demographics. 
However, both datasets adopt the Amazon Mechanical Turk for annotation, which primarily consists of US workers in early stages \citep{ross2009turkers,ipeirotis2010demographics,irani2015difference}}, diversifying the annotation process can include more comprehensive opinions from a broader audience.
Such an annotation can serve either as a training corpus that leads to a better customized model, or as an evaluation set to understand the limitations of the existing models.
This is especially important as UI animation interpretation is a subjective task, therefore leading to diverse annotations \citep{plank-2022-problem, deng-etal-2023-annotate}. 
For example, in stock or financial software, red often indicates an increase in some Asian countries but a decrease in the U.S. 
Incorporating greater cultural diversity among annotators could enrich the dataset and reveal additional insights into cross-cultural differences in how animations and visual cues are interpreted.
When constructing \dataname, we included ten annotations for each animation interpretation, hoping to cover as many cases as possible.
We encourage future efforts in investigating the subjectivity in the task of UI animation understanding and extending the annotations beyond western countries.

Third, in this paper, we try our best to include a comprehensive set of VLMs in our experiments, including nine models from GPT, Gemini, and other model families.
However, as the field is rapidly evolving, it is not feasible to exhaustively evaluate every available model variant.
Another concern is whether to experiment with smaller models.
We have conducted pre-liminary experiments in Appendix \ref{app-sec: model-selection-rationale} and found that smaller models, due to their design constraints (e.g., limited context length, single-image input, etc), cannot handle the UI animation task well.
Therefore, we focus primarily on the nine VLMs in \Cref{tab:VLMs}.
We encourage future efforts from our community to experiment with other VLMs on UI animation understanding.

\section*{Acknowledgments}
This research was funded, in part, by the U.S. Government under ARPA-H contract 1AY2AX000062. The views and conclusions contained in this document are those of the authors and should not be interpreted as representing the official policies, either expressed or implied, of the U.S. Government.

\bibliography{bib}

\appendix
\section{Model Selection Rationale}
\label{app-sec: model-selection-rationale}
The task of UI animation understanding presents unique challenges compared to typical video understanding tasks. 
Unlike standard video understanding where sparse frames could be sufficient \citep{lei-etal-2023-revealing}, perceiving UI animations requires dense frame extraction to capture fine-grained motion. 
This requires models to have comparatively larger context lengths to accommodate longer sequences of frames. 
Additionally, inferring the underlying purpose of an animation requires complex reasoning over interface elements, motion patterns, and context. Therefore, we prioritize larger models with advanced reasoning capabilities and selected state-of-the-art commercial and open-source multimodal large language models listed in \Cref{tab:VLMs}.

\begin{figure}[t]
    \centering
  \includegraphics[width=\linewidth]{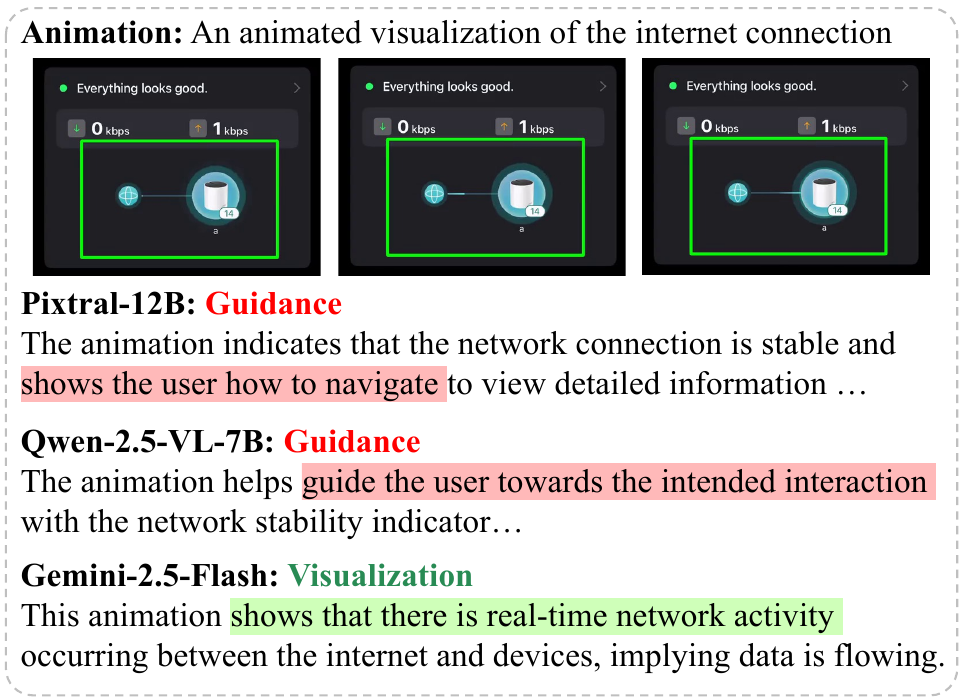}
  \caption{
  An example where small models failed the UI animation understanding task, while the advanced model (Gemini-2.5-Flash) succeeded. 
  The generated interpretation from these small models suggests that these models are not yet capable of robust animation perception and interpretation.
  }
  \label{fig:ap_model_example}
\end{figure}

In addition, we examined smaller VLMs with model sizes ranging from 7B to 14B and observed several limitations that hinder reliable evaluation. 
As a result, we exclude these smaller models from our primary analysis. 
In particular, models with restricted context lengths, such as Qwen2.5-VL 7B Instruct (32k tokens) \citep{bai2025qwen2}, struggle to accommodate a sufficient volume of motion frames for animation interpretation. Other models, such as Llama 3.2 11B Vision Instruct \citep{grattafiori2024llama}, are primarily for image understanding and cannot be fed with multiple images. 
As illustrated in \Cref{fig:ap_model_example}, preliminary tests with smaller models like Pixtral-12B \citep{agrawal2024pixtral} and Qwen-2.5-VL-7B \citep{bai2025qwen2} reveal the failure to perceive animation and its temporal changes, resulting in incorrect categorization and interpretation. 
These preliminary results suggest that existing smaller models are not yet capable of UI animation understanding.
Therefore, we focus primarily on the VLMs listed in \Cref{tab:VLMs}.

All model inference in this work was performed through OpenRouter \url{https://openrouter.ai/}. Model settings were left as default.

\section{Annotation Details}
\label{ap:anno_details}
\begin{figure}[t]
    \centering
  \includegraphics[width=\linewidth]{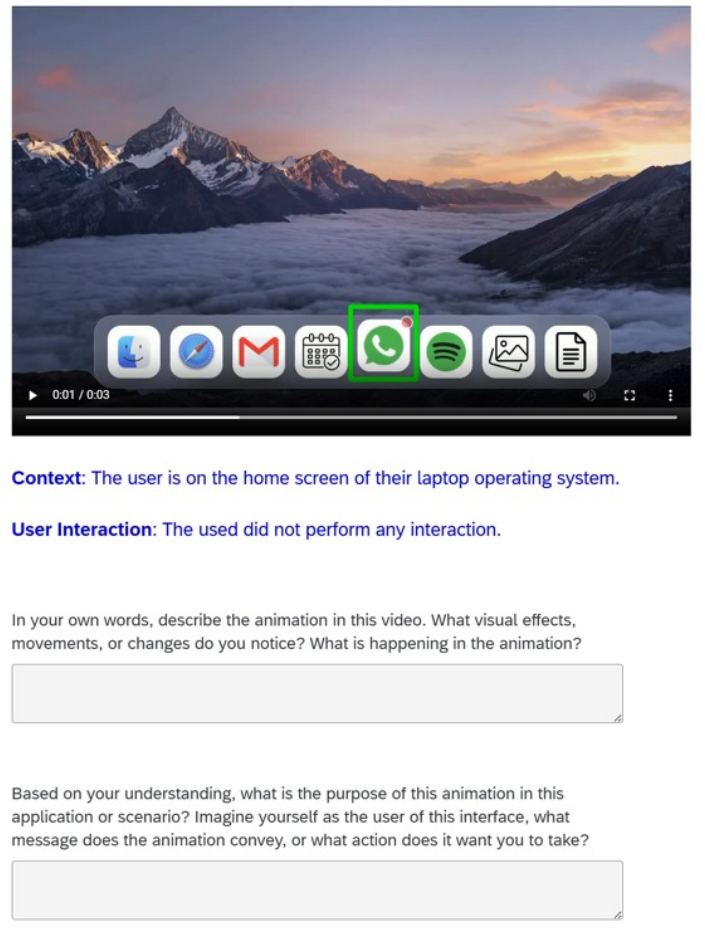}
  \caption{
  An example of the labeling interface, where the annotator can play the animation video with the green bounding box highlighting the animated region, see the context and user interaction details, and provide their interpretations.
  }
  \label{fig:ap_labeling_interface}
\end{figure}
\subsection{Annotation Setup}
We recruited 300 unique participants from Prolific, each of whom annotated a set of 10 videos through a short survey as illustrated in \Cref{fig:ap_labeling_interface}, resulting in a total of 3000 responses. Participants were compensated \$3 per 10 responses. The study is IRB approved, and all participants provided informed consent prior to participation.

The annotation task was hosted on Qualtrics in the form of a short survey, with each session consisting of 10 videos. To start, participants were given a tutorial of the labeling interface, task details, annotation best practices and requirements, and example high-quality annotations. Participants were specifically instructed to focus on the animation within the green bounding box to minimize distractions from other concurrent animations. Participants can modify their answers or revisit the tutorial materials at any time during the session.

\subsection{Ethical Considerations}
To protect participants from exposure to sensitive or potentially harmful content, all videos were manually reviewed by the research team prior to annotation. This verification process ensured that the dataset contained no sensitive material (e.g., sexual content, violence) or potentially harmful visual stimuli (e.g., rapid flashing). Participants were informed of their right to withdraw from the study at any time and were provided with contact information for the research team to address any concerns.

\subsection{Annotator Demographics}
All participants were recruited from within the United States and reported English as their primary language. To increase annotation quality, we recruited participants who have finished at least 1000 tasks on Prolific before, and has an approval rate of 100\%. All annotators were 18 years of age or older, with a mean age of 44.26 years (SD = 13.46). The gender distribution was 158 (52.8\%) female, 140 (46.8\%) male, and 1 (0.3\%) participant who preferred not to disclose their gender. 

\subsection{Annotation Filtering}
To preserve the authenticity of human interpretation, we applied minimal filtering, excluding only empty or inappropriate responses. As a result, the dataset retains brief annotations and explicit expressions of uncertainty (e.g., ``I don't know''). This decision ensures that the data captures the inherent ambiguity of the animations. For example, if a visual stimulus is confusing to human annotators, we expect a robust VLM to reflect similar uncertainty to achieve true alignment.

\subsection{Animation Purpose Categorization}
Animation purposes were annotated by three domain experts. All experts were provided with detailed definitions of each category, along with three example animations per category, to establish a shared understanding of the distinctions between classes. The original annotations exhibited an inter-annotator agreement of Krippendorff’s $\alpha = 0.78$, indicating substantial and reliable agreement despite some disagreements. Final labels were determined by majority voting across annotators. In cases where no majority was reached (i.e., all three annotators assigned different labels), the instances were discussed in a follow-up adjudication session to reach consensus. All experts were compensated at \$20/hr. 

\subsection{Intended Use of the Dataset}
The dataset created in this work is intended for research use, such as evaluation and benchmarking. The dataset is constructed from publicly available content and does not include sensitive information or personally identifiable data. 

\subsection{Dataset Documentation}
\begin{itemize}[noitemsep, topsep=0pt]
    \item Dataset size: 300
    \item Data Coverage: Video recordings of public mobile apps, operating systems, and websites.
    \item Video Language: English
    \item Annotation Language: English
    \item Annotator Region: United States
    \item Annotator Age: average 44.26 (20-80).
    \item Annotator Gender: 52.8\% female, 46.8\% male, 0.3\% undisclosed. 
\end{itemize}

\subsection{Informed Consent}
Below is the informed consent for data annotation:

You are invited to participate in a research study about evaluating machine learning model's understanding of animations used in user interfaces (UI), such as mobile apps, desktop software, and web interfaces. Specifically, the project investigates whether these models can perceive, interpret, and understand user interfaces the same way as humans do. To answer this question, researchers will evaluate human understanding of various UI examples, and then compare the results with machine learning model’s responses to the same questions and see to what extent two sets of answers align with each other.
If you agree to be part of the research study, you will be asked to watch recordings of UI animations and provide your interpretations of them. You will annotate 10 examples. You are not required to finish all examples, and can end the study anytime. We will primarily collect data through your responses in the questionnaire. We will protect the confidentiality of your research records by storing data on a secure server. We do not collect your identifiable information (e.g., your name, email).
There is no direct personal benefit from being in this study. The risks and discomfort associated with participation in this study are minimal.
Compensation: You will receive \$3 for finishing annotating 10 samples. 
Participating in this study is completely voluntary.  Even if you decide to participate now, you may change your mind and stop at any time.  You may choose not to watch any UI recordings, interact with the labeling interface, answer any survey question, or continue with the study for any reason.
If you have questions about this research study, please contact the researcher. 
If you agree to participate, please proceed to the study below.

\section{RQ1 Evaluation Setup}
\label{ap:rq1_primitive}
We created a 3-second clip at 60 fps for each animation effect and used these clips in the RQ1 evaluation. Each prompt was repeated ten times, with answer choices randomly permuted to mitigate potential biases due to option ordering \citep{zheng2024large}. For each run, the same randomized ordering was used across all models to ensure fair and consistent comparisons.
\begin{quote}
\begin{adjustwidth}{-1em}{-1em} 
\begingroup
\small\ttfamily\raggedright

You are given a sequence of frames, uniformly sampled at 10 frames per second from a video of an animation.\\[0.5em]

\textbf{Task:}\\
Identify which single animation type best matches the video you observe.\\[0.5em]

\textbf{Options:}
\begin{enumerate}[label=\Alph*., leftmargin=*, nosep]
  \item Move (object moves in any direction)
  \item Rotate (object rotates along any axis)
  \item Size (object changes sizes along any axis)
  \item Color (object changes in hue, saturation, or brightness)
  \item Fade (object change in transparency/opacity)
  \item Blur (object change in sharpness or clarity)
  \item Morph (object transformation from one shape/form to another)
\end{enumerate}

\ttfamily
\textbf{Output format:}\\
First line: the single letter (A to G) that corresponds to the animation type.
Second line: an explanation of why this animation type matches the video.

\endgroup
\end{adjustwidth}
\end{quote}

\section{RQ2 Evaluation Setup}
\label{ap:rq2_purposes}
\subsection{Definition of Animation Purposes}
Our animation categorization and definitions are derived from prior literature on UI animation taxonomy \cite{DanielLiddle2016, Betrancourt2000, FannyChevalier2016Ani25}, including:

\begin{description}[leftmargin=0pt, labelsep=0.1em, itemsep=0pt]
  \item[\textbf{Transition (Transit.):}] Animations that support layout changes.\\
  Example: A flame animation in a privacy browser burning away tabs to transition to a new session.

{\centering
  \includegraphics[width=0.9\linewidth]{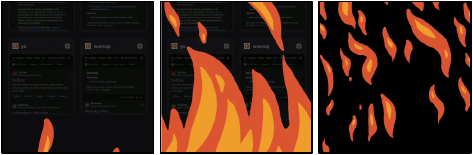}\par
}
  \item[\textbf{Demonstration (Demo.):}] Animations that reveal or explain the behavior, functionality, or structure of the interface and its elements.\\
  Example: An animation of the Face ID setup demo.

{\centering
  \includegraphics[width=0.9\linewidth]{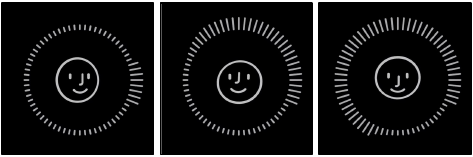}\par
}
  \item[\textbf{Guidance (Guide):}] Animations that guide the user towards an intended interaction.\\
  Example: An animated arrow guiding the user to swipe up to capture the Pokemon. 

{\centering
  \includegraphics[width=0.9\linewidth]{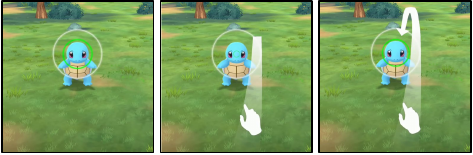}\par
}
  \item[\textbf{Feedback:}] Animations that provide visual responses to user interactions.\\
  Example: A ripple animation appears when two iPhones are near each other for proximity AirDrop. 

{\centering
  \includegraphics[width=0.9\linewidth]{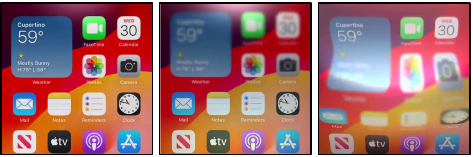}\par
}
  \item[\textbf{Visualization (Vis.):}] Animations that represent system status, data, or other information.\\
  Example: An animated bottle icon gradually filling up to visualize the loading process.

{\centering
  \includegraphics[width=0.9\linewidth]{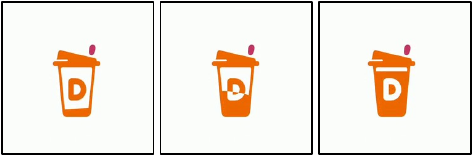}\par
}
  \item[\textbf{Highlight:}] Animations that emphasize specific content or draw the user's attention to key elements.\\
  Example: A pulsing ripple animation highlighting the menu button in the corner. 

{\centering
  \includegraphics[width=0.9\linewidth]{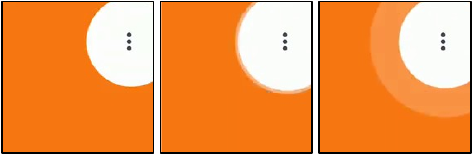}\par
}
  \item[\textbf{Aesthetic:}] Animations that enhance the visual appeal, create an emotional impact, or improve user experiences.\\
  Example: Animated confetti falling from the top. 

{\centering
  \includegraphics[width=0.9\linewidth]{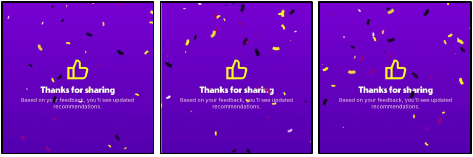}\par
}
\end{description}

\subsection{Evaluation Prompt}
\begin{quote}
\begin{adjustwidth}{-1em}{-1em} 
\begingroup
\small\ttfamily\raggedright
You are a UI animation expert. You will analyze an ordered sequence of frames sampled uniformly at 10 fps from a user-interface (UI) animation. Within each video, a green box will appear when the animation starts, and disappear when the animation ends. Please primarily focus on the animation happening within the green box when you answer the questions. Please see all the frames, and answer the following questions about the UI animation in this video. \\

You will be given the following information as Inputs\\
- frames: a sequence of images captured at 10 fps. A green box will appear to identify the region of animation.\\
- context: brief description of the situation (e.g., app, user goal)\\
- input: description of any user interaction right before or during the animation (tap, swipe, talk, etc.), or no input was actively performed.\\

\textbf{Data for this video}\\
context: \{context\}\\
input: \{input\}\\

Question: What is the primary purpose of this UI animation? Describe your rationale and explain how the animation effect supports that purpose. Single-answer question. Select only one option. \\

\textbf{Options:}
\begin{enumerate}[label=\Alph*., leftmargin=*, nosep]
\item Transition: Animations that support layout changes.
\item Demonstration: Animations that reveal or explain the behavior, functionality, or structure of the interface and its elements.
\item Guidance: Animations that guide the user towards an intended interaction
\item Feedback: Animations that provide visual responses to user interactions.
\item Visualization: Animations that represent system status, data, or other information.
\item Highlight: Animations that emphasize specific content or draw the user's attention to key elements.
\item Aesthetic: Animations that enhance the visual appeal, create an emotional impact, or improve user experiences.
\end{enumerate}

For the selected category, write a sentence describing your rationale and explain how the animation effect supports that purpose

\textbf{Output format:}\\
Write exactly one line for the selected category and its explanation/description. For example:
<Letter> — <PurposeName>: <Your rationale>

\endgroup
\end{adjustwidth}
\end{quote}

\section{RQ2 Additional Results}
\label{app-subsec:rq2-additional-discussion}

\subsection{Error Patterns and Category Confusions}

\begin{figure}[t]
  \centering
  \includegraphics[width=0.9\linewidth]{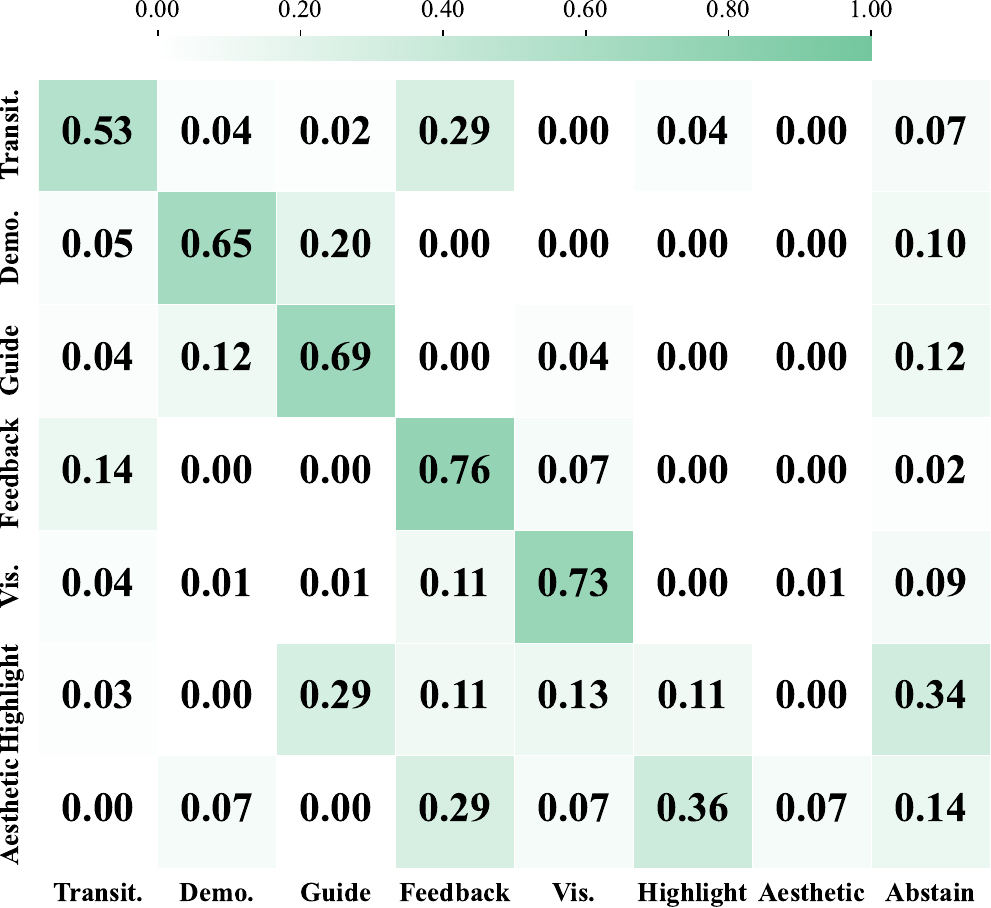}
  \caption{Accuracy of the majority vote of predictions from nine models. 
  The y-axis denotes ground-truth labels, and the x-axis denotes majority-vote predictions across models. 
  Videos without a majority (fewer than five agreeing models) are labeled as ``Abstain.''
  Transition, Demonstration, Guidance, Feedback, and Visualization show the strongest performance and consistency, whereas Highlight and Aesthetic exhibit the weakest.
  }
  \label{fig:q2_confusion_matrix}
\end{figure}
\begin{figure*}[t]
  \centering
  \includegraphics[width=\linewidth]{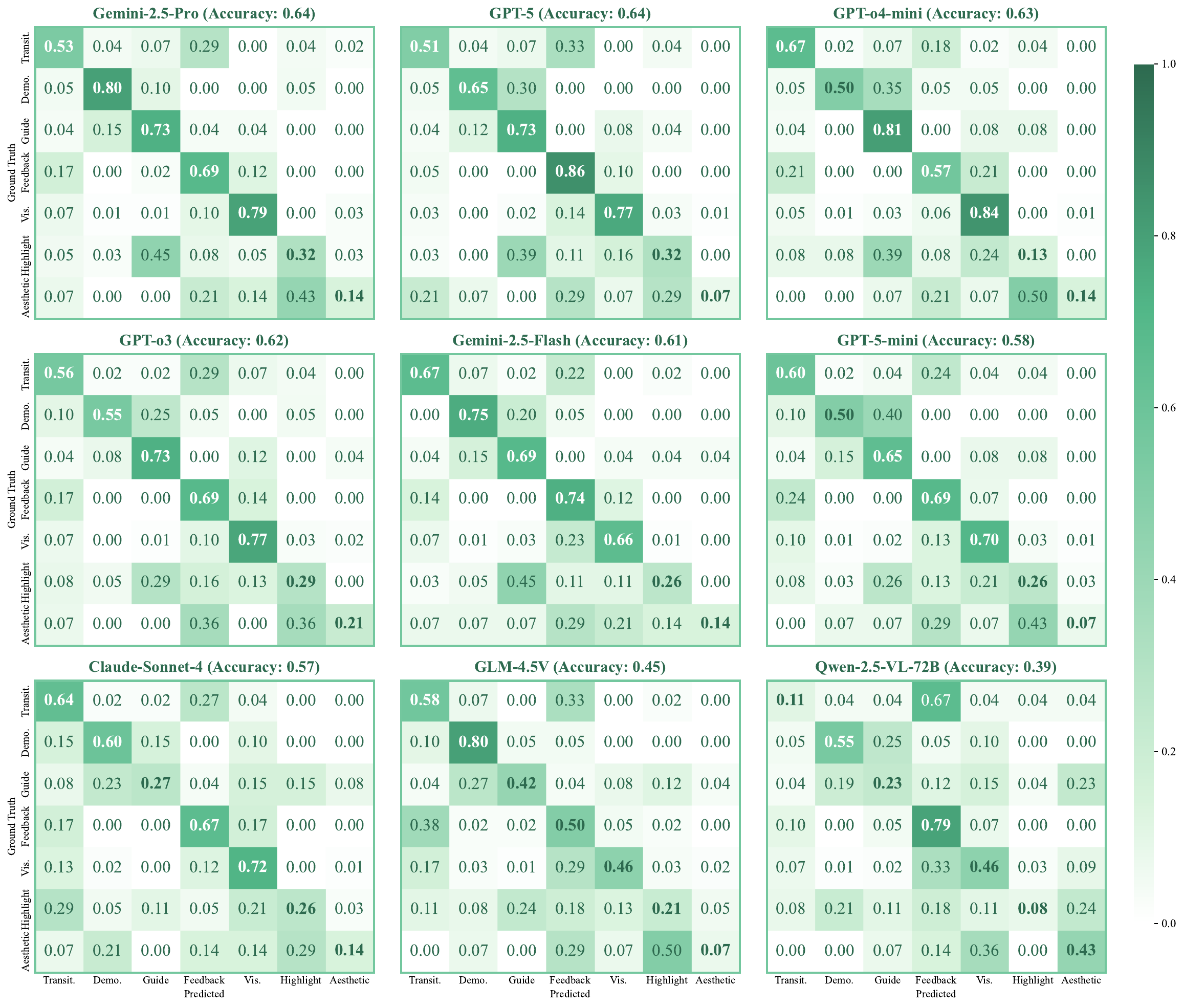}
  \caption{Confusion matrix for each model. }
  \label{fig:q2_all_confusion_matrices}
\end{figure*}
\Cref{fig:q2_all_confusion_matrices} shows the confusion matrix of individual models, and \Cref{fig:q2_confusion_matrix} shows the confusion matrix of the majority-voted predictions. For majority-vote predictions across models, when there are fewer than 5 models agreeing on the same answer, the prediction will be labeled as ``abstain''. This is to reflect an overview of how VLMs in general performs on the categorization task. 
A closer inspection of misclassified cases reveals distinct error patterns. 
Animations in the Highlight category frequently fail to reach a majority consensus (i.e., Abstain; 13 out of 38 cases) or are misclassified as Guidance (11 out of 38). 
Similarly, Aesthetic animations are most often misclassified as Feedback (5 out of 14) or Highlight (5 out of 14). 
In addition, \autoref{fig:q2_confusion_matrix} reveals several bidirectional confusion pairs, including Transition and Feedback (T as F: 13, F as T: 6), Feedback and Visualization (F as V: 3, V as F: 13), and Demonstration and Guidance (D as G: 4, G as D: 3). 
These confusions occur primarily between conceptually adjacent categories, suggesting that models struggle to capture fine-grained distinctions between closely related animation purposes, both visually and conceptually. Additionally, the systematically lower performance for Highlight and Aesthetic categories suggests that models are less effective at recognizing animations that serve subtle affective or cognitive purposes, indicating that the conceptual categories may be ``memorized'' than cognitively ``perceived'' by VLMs. This shows that a cognitive gap still exist between VLMs and human users in understanding these subtle interface cues. 

\subsection{Performance on Video Models}
We conduct additional evaluations using Video LLaMa, LLaVA-Video, Qwen-2.5-VL, and Gemini-2.5-pro, where we use videos directly as input. The results are listed below:
\begin{table}[h]
\centering
\small
\setlength{\tabcolsep}{4pt}
\begin{tabular}{lcc}
\toprule
Model & Acc & MacF1 \\
\midrule
gemini-2.5-pro           & 0.63 & 0.55 \\
qwen2.5-vl-72b-instruct  & 0.42 & 0.38 \\
videollama3-7b-local     & 0.21 & 0.22 \\
llava-video-7b-local     & 0.19 & 0.20 \\
\bottomrule
\end{tabular}
\caption{Model performance with video input}
\label{tab:model_performance}
\end{table}

\subsection{Statistical Significance Test}
\label{app-subsec: rq2-stats-test}
\paragraph{Test selection.}
We use McNemar's test \citep{mcnemar1947note} to compare system variants, as our evaluation is paired and yields binary correctness outcomes for each video instance. 
Specifically, each system configuration produces a categorical prediction for the same set of videos, which we evaluate against the human-annotated ground truth to determine whether the prediction is correct or incorrect. 
McNemar’s test assesses whether two classifiers differ significantly when tested on the same examples, while accounting for the dependency between paired observations. 

\paragraph{Test results.} 
We conducted McNemar's test between each pair of VLMs in \Cref{tab:rq2_model_performance} to investigate whether their performance is statistically different. \Cref{tab:rq2-stats-test} shows the result.
We highlight that more than half of the pairs yield a difference that is statistically significant with $p<0.05$ or $p<0.1$.

\begin{table}[t]
\small
\centering
\resizebox{\linewidth}{!}{
\begin{tabular}{llrcc}
\toprule
Model 1          & Model 2          & \multicolumn{1}{l}{p value} & \multicolumn{1}{l}{p < 0.05} & \multicolumn{1}{l}{p < 0.1} \\
\midrule
Gemini-2.5-Pro   & Qwen2.5-VL-72B   & 4.25e-13                    & True                         & True                        \\
GPT-5            & Qwen2.5-VL-72B   & 1.20e-12                    & True                         & True                        \\
GPT-o3           & Qwen2.5-VL-72B   & 5.56e-11                    & True                         & True                        \\
GPT-o4-mini      & Qwen2.5-VL-72B   & 7.47e-11                    & True                         & True                        \\
Gemini-2.5-Flash & Qwen2.5-VL-72B   & 5.68e-10                    & True                         & True                        \\
GPT-5-mini       & Qwen2.5-VL-72B   & 2.87e-09                    & True                         & True                        \\
Gemini-2.5-Pro   & GLM-4.5V         & 2.08e-08                    & True                         & True                        \\
GLM-4.5V         & GPT-5            & 3.22e-08                    & True                         & True                        \\
Claude-sonnet-4  & Qwen2.5-VL-72B   & 7.68e-08                    & True                         & True                        \\
GLM-4.5V         & GPT-o4-mini      & 2.88e-07                    & True                         & True                        \\
GLM-4.5V         & GPT-o3           & 3.72e-07                    & True                         & True                        \\
Gemini-2.5-Flash & GLM-4.5V         & 4.75e-07                    & True                         & True                        \\
GLM-4.5V         & GPT-5-mini       & 1.30e-05                    & True                         & True                        \\
Claude-sonnet-4  & GLM-4.5V         & 2.24e-04                    & True                         & True                        \\
Claude-sonnet-4  & GPT-5            & 2.57e-02                    & True                         & True                        \\
Claude-sonnet-4  & Gemini-2.5-Pro   & 2.63e-02                    & True                         & True                        \\
GPT-5            & GPT-5-mini       & 3.31e-02                    & True                         & True                        \\
Claude-sonnet-4  & GPT-o4-mini      & 3.56e-02                    & True                         & True                        \\
Gemini-2.5-Pro   & GPT-5-mini       & 4.74e-02                    & True                         & True                        \\
GLM-4.5V         & Qwen2.5-VL-72B   & 6.71e-02                    & False                        & True                        \\
Claude-sonnet-4  & GPT-o3           & 8.05e-02                    & False                        & True                        \\
GPT-5-mini       & GPT-o4-mini      & 9.80e-02                    & False                        & True                        \\
GPT-5-mini       & GPT-o3           & 1.48e-01                    & False                        & False                       \\
Claude-sonnet-4  & Gemini-2.5-Flash & 2.42e-01                    & False                        & False                       \\
Gemini-2.5-Flash & GPT-5            & 2.60e-01                    & False                        & False                       \\
Gemini-2.5-Flash & Gemini-2.5-Pro   & 2.66e-01                    & False                        & False                       \\
Gemini-2.5-Flash & GPT-5-mini       & 4.64e-01                    & False                        & False                       \\
Gemini-2.5-Flash & GPT-o4-mini      & 4.89e-01                    & False                        & False                       \\
GPT-5            & GPT-o3           & 5.33e-01                    & False                        & False                       \\
Gemini-2.5-Pro   & GPT-o3           & 5.56e-01                    & False                        & False                       \\
Gemini-2.5-Flash & GPT-o3           & 6.25e-01                    & False                        & False                       \\
Claude-sonnet-4  & GPT-5-mini       & 7.16e-01                    &   False                           & False                       \\
Gemini-2.5-Pro   & GPT-o4-mini      & 7.24e-01                    & False                        & False                       \\
GPT-5            & GPT-o4-mini      & 7.98e-01                    & False                        & False                       \\
GPT-o3           & GPT-o4-mini      & 8.90e-01                    & False                        & False                       \\
Gemini-2.5-Pro   & GPT-5            & 1.00                        & False                        & False                             \\
\bottomrule
\end{tabular}}
\caption{Pair-wise statistical significance test for results reported in \Cref{tab:rq2_model_performance}.
We highlight that more than half of the pairs yield a difference that is statistically significant with $p<0.05$ or $p<0.1$.
}
\label{tab:rq2-stats-test}
\end{table}

\section{RQ3 Evaluation Setup}
\label{ap:rq3_alignment}
\subsection{Comparison Methods}
\label{app-subsec: rq3-eval-methods}
\begin{figure}[t]
    \centering
  \includegraphics[width=\linewidth]{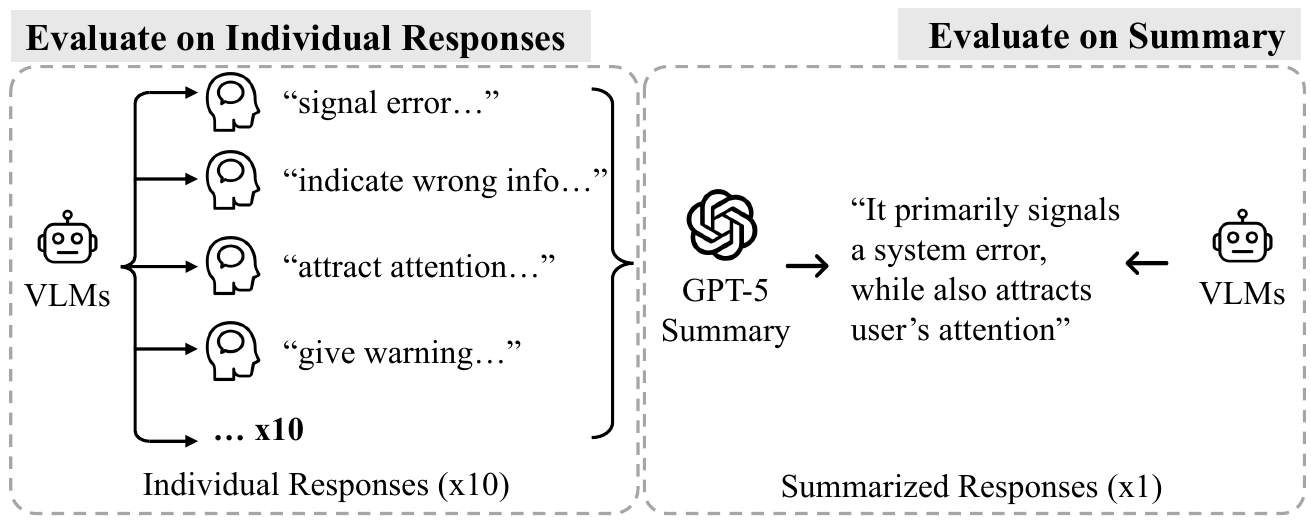}
  \caption{
  Illustration of semantic similarity computed against individual annotations ($N=10$) versus a consolidated summary ($N=1$).
  }
  \label{fig:ap_rq3_method}
\end{figure}

As shown in \Cref{fig:ap_rq3_method}, we conducted two types of comparisons. The first approach (\Cref{fig:ap_rq3_method} left) compares the VLM interpretation with each of the 10 human interpretations for a specific video, resulting in 10 similarity scores per video. The second approach (\Cref{fig:ap_rq3_method} right) uses GPT-5 to summarize all 10 interpretations into a single response and compares the VLM interpretation with the summarized response, resulting in one score. These two methods offer perspectives at different levels of granularity. While the individual comparisons are susceptible to variations in annotation quality (e.g., short or unclear responses), they capture the distribution of direct similarities. Conversely, the summarized approach reflects alignment with the overall human understanding and focuses on high-level concepts, though it may lose some specific details found in individual responses. Empirically, we found that both approaches yield similar model rankings. Therefore, we report the results of the summarized responses in the main text and provide the individual comparison results in Appendix \ref{app-subsec: rq3-additional-results} for additional context.

\subsection{Evaluation Prompt}
\label{app-subsec: rq3-eval-prompt}
\begin{quote}
\begin{adjustwidth}{-1em}{-1em} 
\begingroup
\small\ttfamily\raggedright
Please act as an impartial judge and compare two short texts (Text A and Text B) that describe the purpose/interpretation of the same UI animation. Decide their semantic equivalence and coverage, considering:
\begin{itemize}
    \item Topics and actions, entities, and roles
    \item Key attributes: numbers, units, dates/times, polarity/negation
    \item Causal/temporal relations and constraints
\end{itemize}

\textbf{Scoring (choose exactly one numeric score):}
\begin{itemize}
    \item[5:] Paraphrase/equivalent meaning — Fully equivalent or one fully contains the other with no contradictions. No missing key facts.
    \item[4:] Nearly equivalent; minor nuance differences — Main points identical, only subtle wording or emphasis differences.
    \item[3:] Same gist; missing/extra key detail(s) — Core idea matches but some important details missing, added, or slightly inconsistent.
    \item[2:] Some overlap; key differences — Partial overlap in main topic but significant differences in specifics or interpretation.
    \item[1:] Same topic only — Related to same general subject but different focus, purpose, or approach.
    \item[0:] Unrelated or contradictory — Completely unrelated topics or directly contradictory statements.
\end{itemize}

\textbf{Output Format:}
Return STRICT JSON (no code fences) with schema:
\begin{verbatim}
{"score": 5 | 4 | 3 | 2 | 1 | 0, "reason": "..."}
\end{verbatim}

Be concise and objective. Avoid any position biases and ensure that the order in which the responses were presented does not influence your decision. Do not allow the length of the responses to influence your evaluation. Be as objective as possible.

Text A: \{text\_a\}

Text B: \{text\_b\}
\endgroup
\end{adjustwidth}
\end{quote}

\section{RQ3 Additional Results}
\label{ap:rq3-additional-results}
\subsection{Individually Compared Results}
\label{app-subsec: rq3-additional-results}

\Cref{tab:rq3_vlm_individual_result} shows the similarity score statistics if individually compared with human annotation. 
Comparing with individual (\Cref{tab:rq3_vlm_individual_result}) or summarized (\Cref{tab:rq3_vlm_result}) response both yield similar model rankings.

\subsection{Statistical Significance Test}
\label{app-subsec:rq3-stats-test}

\begin{table}[t]
\resizebox{\linewidth}{!}{
\begin{tabular}{llrcc}
\toprule
Model 1          & Model 2          & \multicolumn{1}{l}{p value} & \multicolumn{1}{l}{p < 0.05} & \multicolumn{1}{l}{p < 0.1} \\
\midrule
GLM-4.5V         & GPT-o3           & 1.56e-13                    & True                         & True                        \\
GLM-4.5V         & GPT-5            & 3.49e-13                    & True                         & True                        \\
GLM-4.5V         & GPT-5-mini       & 1.42e-11                    & True                         & True                        \\
Gemini-2.5-Pro   & GLM-4.5V         & 8.33e-11                    & True                         & True                        \\
GPT-o3           & Qwen2.5-VL-72B   & 2.15e-10                    & True                         & True                        \\
GPT-5-mini       & Qwen2.5-VL-72B   & 2.75e-09                    & True                         & True                        \\
Gemini-2.5-Flash & GLM-4.5V         & 1.26e-08                    & True                         & True                        \\
GPT-5            & Qwen2.5-VL-72B   & 3.98e-08                    & True                         & True                        \\
GLM-4.5V         & GPT-o4-mini      & 1.00e-07                    & True                         & True                        \\
Gemini-2.5-Pro   & Qwen2.5-VL-72B   & 3.30e-07                    & True                         & True                        \\
Claude-Sonnet-4  & GPT-o3           & 1.30e-06                    & True                         & True                        \\
Claude-Sonnet-4  & GPT-5            & 7.10e-06                    & True                         & True                        \\
Gemini-2.5-Flash & Qwen2.5-VL-72B   & 3.49e-05                    & True                         & True                        \\
Claude-Sonnet-4  & GPT-5-mini       & 5.08e-05                    & True                         & True                        \\
GPT-o3           & GPT-o4-mini      & 7.90e-05                    & True                         & True                        \\
Claude-Sonnet-4  & GLM-4.5V         & 1.23e-04                    & True                         & True                        \\
Claude-Sonnet-4  & Gemini-2.5-Pro   & 1.48e-04                    & True                         & True                        \\
GPT-o4-mini      & Qwen2.5-VL-72B   & 2.84e-04                    & True                         & True                        \\
GPT-5            & GPT-o4-mini      & 9.31e-04                    & True                         & True                        \\
Claude-Sonnet-4  & Gemini-2.5-Flash & 3.46e-03                    & True                         & True                        \\
GPT-5-mini       & GPT-o4-mini      & 4.52e-03                    & True                         & True                        \\
Gemini-2.5-Flash & GPT-o3           & 1.50e-02                    & True                         & True                        \\
Gemini-2.5-Pro   & GPT-o4-mini      & 1.78e-02                    & True                         & True                        \\
GLM-4.5V         & Qwen2.5-VL-72B   & 4.38e-02                    & True                         & True                        \\
Claude-Sonnet-4  & Qwen2.5-VL-72B   & 6.93e-02                    & False                        & True                        \\
GPT-5-mini       & GPT-o3           & 8.59e-02                    & False                        & True                        \\
Gemini-2.5-Flash & GPT-5            & 8.85e-02                    & False                        & True                        \\
Claude-Sonnet-4  & GPT-o4-mini      & 1.07e-01                    & False                        & False                       \\
Gemini-2.5-Flash & Gemini-2.5-Pro   & 1.71e-01                    & False                        & False                       \\
Gemini-2.5-Flash & GPT-5-mini       & 1.79e-01                    & False                        & False                       \\
Gemini-2.5-Pro   & GPT-o3           & 1.87e-01                    & False                        & False                       \\
Gemini-2.5-Flash & GPT-o4-mini      & 2.49e-01                    & False                        & False                       \\
GPT-5            & GPT-5-mini       & 3.97e-01                    & False                        & False                       \\
Gemini-2.5-Pro   & GPT-5            & 4.68e-01                    & False                        & False                       \\
GPT-5            & GPT-o3           & 6.23e-01                    & False                        & False                       \\
Gemini-2.5-Pro   & GPT-5-mini       & 7.99e-01                    & False                        & False     \\
\bottomrule
\end{tabular}}
\caption{Pair-wise statistical significance test for results reported in \Cref{tab:rq3_vlm_result}.
We highlight that most pairs yield a difference that is statistically significant with $p<0.05$ or $p<0.1$.
}
\label{tab:rq3-stats-test}
\end{table}

\paragraph{Test selection.}

To compare scores of the model-generated UI animation interpretation, we use the Wilcoxon signed-rank test \citep{wilcoxon1945individual}.
In this setting, the scores for the same set of video instances yield paired but non-normally distributed observations. 
The Wilcoxon signed-rank test makes no assumptions about score normality, respects the paired structure of the evaluation, and tests whether the median difference between two systems’ scores is zero. 
We therefore adopt the Wilcoxon signed-rank test to assess pairwise performance differences in \Cref{tab:rq3-stats-test}.

\paragraph{Test results.}
We conducted the Wilcoxon signed-rank test between each pair of VLMs in \Cref{tab:rq3_vlm_result}.
We highlight that most pairs yield a difference that is statistically significant with $p<0.05$ or $p<0.1$.

\begin{table}[t]
\small
\centering
\renewcommand*{\arraystretch}{1.3}
\resizebox{\linewidth}{!}{
\begin{tabular}{lccc}
\toprule
                        & \multicolumn{1}{l}{Mean ($\uparrow$)} & \multicolumn{1}{l}{Std ($\downarrow$)} & \multicolumn{1}{c}{Distribution} \\
\midrule
\raisebox{-0.5ex}{\includegraphics[height=1.2em]{assets/logos/gpt.png}}\,GPT-5-mini
  & \cellcolor[HTML]{68C296}2.76
  & \cellcolor[HTML]{6AC297}1.10
  & \raisebox{-0.4ex}{\includegraphics[height=1.5em]{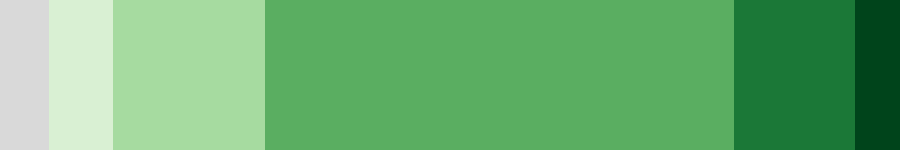}} \\

\raisebox{-0.5ex}{\includegraphics[height=1.2em]{assets/logos/gpt.png}}\,GPT-5
  & \cellcolor[HTML]{70C69C}2.76
  & \cellcolor[HTML]{6AC297}1.14
  & \raisebox{-0.4ex}{\includegraphics[height=1.5em]{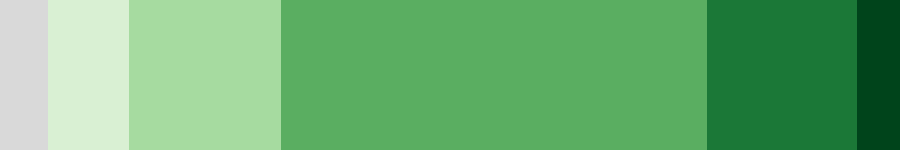}} \\

\raisebox{-0.5ex}{\includegraphics[height=1.2em]{assets/logos/gemini.png}}\,Gemini-2.5-Flash
  & \cellcolor[HTML]{83CDA9}2.74
  & \cellcolor[HTML]{8DD1B0}1.19
  & \raisebox{-0.4ex}{\includegraphics[height=1.5em]{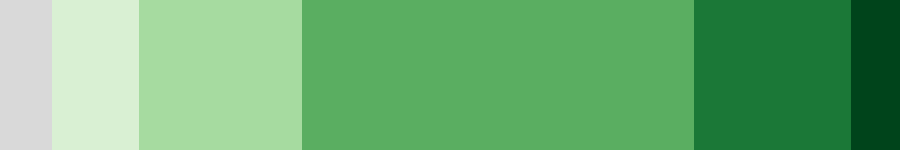}} \\

\raisebox{-0.5ex}{\includegraphics[height=1.2em]{assets/logos/gpt.png}}\,GPT-o3
  & \cellcolor[HTML]{94D4B4}2.73
  & \cellcolor[HTML]{6AC297}1.14
  & \raisebox{-0.4ex}{\includegraphics[height=1.5em]{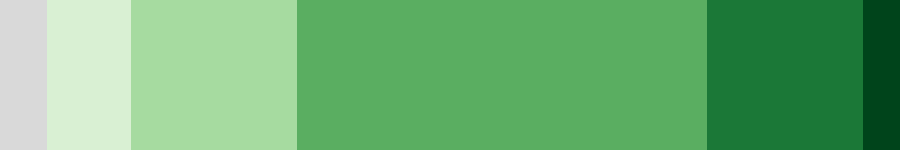}} \\

\raisebox{-0.5ex}{\includegraphics[height=1.2em]{assets/logos/gemini.png}}\,Gemini-2.5-Pro
  & \cellcolor[HTML]{AFDFC7}2.71
  & \cellcolor[HTML]{8DD1B0}1.17
  & \raisebox{-0.4ex}{\includegraphics[height=1.5em]{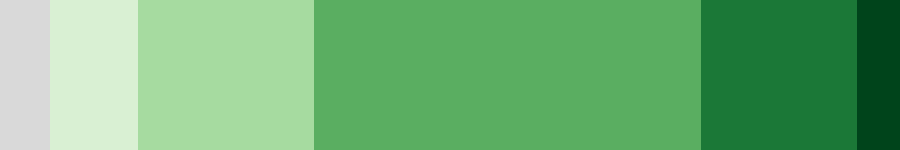}} \\

\raisebox{-0.5ex}{\includegraphics[height=1.2em]{assets/logos/gpt.png}}\,GPT-o4-mini
  & \cellcolor[HTML]{D0ECDE}2.65
  & \cellcolor[HTML]{6AC297}1.14
  & \raisebox{-0.4ex}{\includegraphics[height=1.5em]{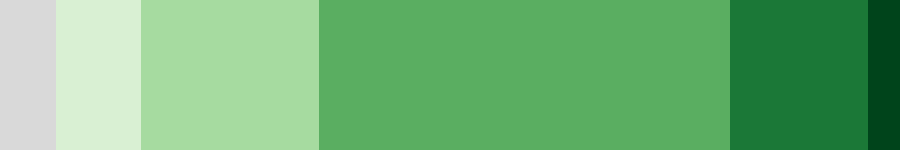}} \\

\raisebox{-0.5ex}{\includegraphics[height=1.2em]{assets/logos/claude.png}}\,Claude-Sonnet-4
  & \cellcolor[HTML]{E6F5EE}2.53
  & \cellcolor[HTML]{B4E0CB}1.26
  & \raisebox{-0.4ex}{\includegraphics[height=1.5em]{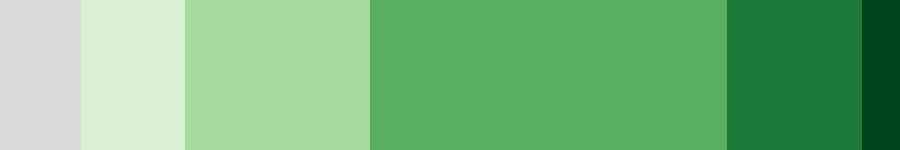}} \\

\raisebox{-0.5ex}{\includegraphics[height=1.2em]{assets/logos/qwen.png}}\,Qwen2.5-VL-72B-Instruct
  & \cellcolor[HTML]{F2FAF7}2.48
  & \cellcolor[HTML]{B4E0CB}1.21
  & \raisebox{-0.4ex}{\includegraphics[height=1.5em]{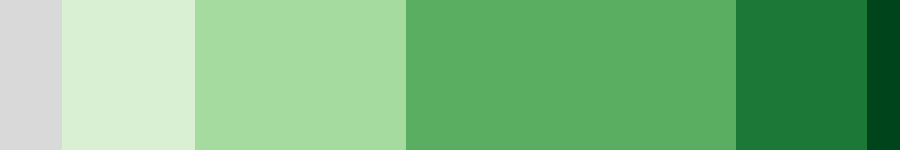}} \\

\raisebox{-0.5ex}{\includegraphics[height=1.2em]{assets/logos/glm.png}}\,GLM-4.5V
  & \cellcolor[HTML]{FFFFFF}2.29
  & \cellcolor[HTML]{FFFFFF}1.27
  & \raisebox{-0.4ex}{\includegraphics[height=1.5em]{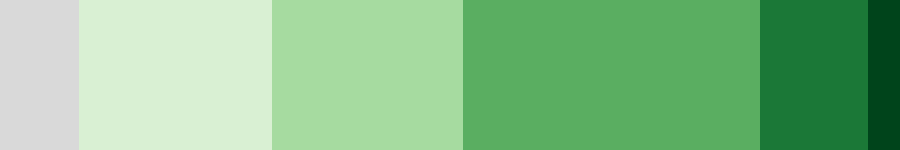}} \\

\bottomrule
\end{tabular}}
\caption{Statistics for semantic similarity scores.
We calculate the score by comparing the model prediction with each individual human's response and report the average score.
Similar to \Cref{tab:rq3_vlm_result}, we report the score distribution, where the five colors from left to right correspond to scores from 0 to 5.}
\label{tab:rq3_vlm_individual_result}
\end{table}

\subsection{Discussions}
\label{app-subec: rq3-discussion}
\paragraph{Feedback presents the most challenges to VLMs.}
Category-wise, Feedback animations exhibited the highest rate of unrelated responses, with 66.7\% receiving one or more unrelated predictions.
In contrast, the corresponding rates were substantially lower for other categories: 11.1\% for Transition, 20.0\% for Demonstration, 23.1\% for Guidance, 22.6\% for Visualization, 26.3\% for Highlight, and 21.4\% for Aesthetic.

We thus investigate whether these VLMs struggle with the Feedback category due to conceptual understanding or perceptual limitations.
For example shown in \Cref{fig:rq3_wrong_example} (top), five models produced responses along the lines of ``the system is giving feedback that it is verifying the password'', which shows that although it can still correctly categorize the high-level animation purpose as Feedback, it fails to recognize the shake itself, or recognize the actual meaning of the shake. 
This highlights a limitation of VLMs in detecting rapid or small-scale movements such as shaking, which in turn prevents accurate interpretation of feedback animations.

This limitation also explains the generally weaker performance of VLMs on Feedback animations. Compared to other categories, feedback animations are often shorter in duration and involve less pronounced graphical change, making them especially reliant on detailed motion cues. The frequent misrecognition of shaking movements suggests that VLMs may face challenges in extracting frame-to-frame changes, motion dynamics, or perceiving visual changes as a whole. Addressing these challenges could be an important avenue for future work, both in evaluating perceptual sensitivity and in developing techniques to improve VLM's perception ability. Despite these limitations, VLMs demonstrated certain amount of overlap with human interpretations in most categories, and when they successfully perceived the animation, they were generally able to reason about its purpose within the context. These findings suggest that while perceptual challenges continue to hinder performance in certain cases, especially subtle or motion-dependent animations, VLMs have potential to capture, and align with, human interpretations of UI animations.
\begin{table}[t]
\small
\centering
\begin{tabular}{llrcc}
\toprule
Setting 1 & Setting 2 & \multicolumn{1}{l}{p value} & \multicolumn{1}{l}{p < 0.05} & \multicolumn{1}{l}{p < 0.1} \\
\midrule
-         & C         & 0.02                        & True                         & True                        \\
-         & CP        & 0.09                        & False                        & True                        \\
-         & M         & 0.46                        & False                        & False                       \\
-         & MC        & 0.22                        & False                        & False                       \\
-         & MCP       & 1.00                        & False                        & False                       \\
-         & MP        & 0.06                        & False                        & True                        \\
-         & P         & 0.24                        & False                        & False                       \\
C         & CP        & 0.47                        & False                        & False                       \\
C         & M         & 0.00                        & True                         & True                        \\
C         & MC        & 0.20                        & False                        & False                       \\
C         & MCP       & 0.01                        & True                         & True                        \\
C         & MP        & 0.00                        & True                         & True                        \\
CP        & M         & 0.01                        & True                         & True                        \\
CP        & MC        & 0.64                        & False                        & False                       \\
CP        & MCP       & 0.03                        & True                         & True                        \\
CP        & MP        & 0.00                        & True                         & True                        \\
M         & MC        & 0.05                        & True                         & True                        \\
M         & MCP       & 0.45                        & False                        & False                       \\
M         & MP        & 0.32                        & False                        & False                       \\
MC        & MCP       & 0.17                        & False                        & False                       \\
MP        & MC        & 0.00                        & True                         & True                        \\
MP        & MCP       & 0.09                        & False                        & True                        \\
P         & C         & 0.00                        & True                         & True                        \\
P         & CP        & 0.01                        & True                         & True                        \\
P         & M         & 0.78                        & False                        & False                       \\
P         & MC        & 0.02                        & True                         & True                        \\
P         & MCP       & 0.27                        & False                        & False                       \\
P         & MP        & 0.64                        & False                        & False                                            \\
          \bottomrule
\end{tabular}
\caption{Pair-wise statistical significance test for purpose categorization (RQ2) results reported in \Cref{tab:factor_ablation}.
``-'' indicates the vanilla model (the base setting in \Cref{tab:factor_ablation}.}
\label{tab: rq4_rq2_stats_test}
\end{table}

\begin{table}[t]
\centering
\small
\begin{tabular}{llrcc}
\toprule
Setting 1 & Setting 2 & \multicolumn{1}{l}{p value} & \multicolumn{1}{l}{p < 0.05} & \multicolumn{1}{l}{p < 0.1} \\
\midrule
-         & C         & 3.43e-02                    & True                         & True                        \\
-         & CP        & 1.38e-05                    & True                         & True                        \\
-         & M         & 2.98e-01                    & False                        & False                       \\
-         & MC        & 1.26e-02                    & True                         & True                        \\
-         & MCP       & 4.59e-07                    & True                         & True                        \\
-         & MP        & 3.08e-05                    & True                         & True                        \\
-         & P         & 1.81e-06                    & True                         & True                        \\
C         & CP        & 4.37e-03                    & True                         & True                        \\
C         & M         & 1.31e-03                    & True                         & True                        \\
C         & MC        & 6.15e-01                    & False                        & False                       \\
C         & MCP       & 1.97e-03                    & True                         & True                        \\
C         & MP        & 6.96e-03                    & True                         & True                        \\
CP        & M         & 5.75e-08                    & True                         & True                        \\
CP        & MC        & 2.30e-02                    & True                         & True                        \\
CP        & MCP       & 3.86e-01                    & False                        & False                       \\
CP        & MP        & 9.17e-01                    & False                        & False                       \\
M         & MC        & 7.44e-04                    & True                         & True                        \\
M         & MCP       & 1.05e-09                    & True                         & True                        \\
M         & MP        & 1.02e-07                    & True                         & True                        \\
MC        & MCP       & 6.28e-03                    & True                         & True                        \\
MP        & MC        & 1.93e-02                    & True                         & True                        \\
MP        & MCP       & 5.05e-01                    & False                        & False                       \\
P         & C         & 2.65e-03                    & True                         & True                        \\
P         & CP        & 5.84e-01                    & False                        & False                       \\
P         & M         & 2.20e-08                    & True                         & True                        \\
P         & MC        & 1.55e-02                    & True                         & True                        \\
P         & MCP       & 9.93e-01                    & False                        & False                       \\
P         & MP        & 7.90e-01                    & False                        & False                    \\
\bottomrule
\end{tabular}
\caption{Pair-wise statistical significance test for UI animation interpretation (RQ3) results reported in \Cref{tab:factor_ablation}.
``-'' indicates the vanilla model (the base setting in \Cref{tab:factor_ablation}.}
\label{tab: rq4_rq3_stats_test}
\end{table}

\begin{figure}[h]
  \includegraphics[width=\linewidth]{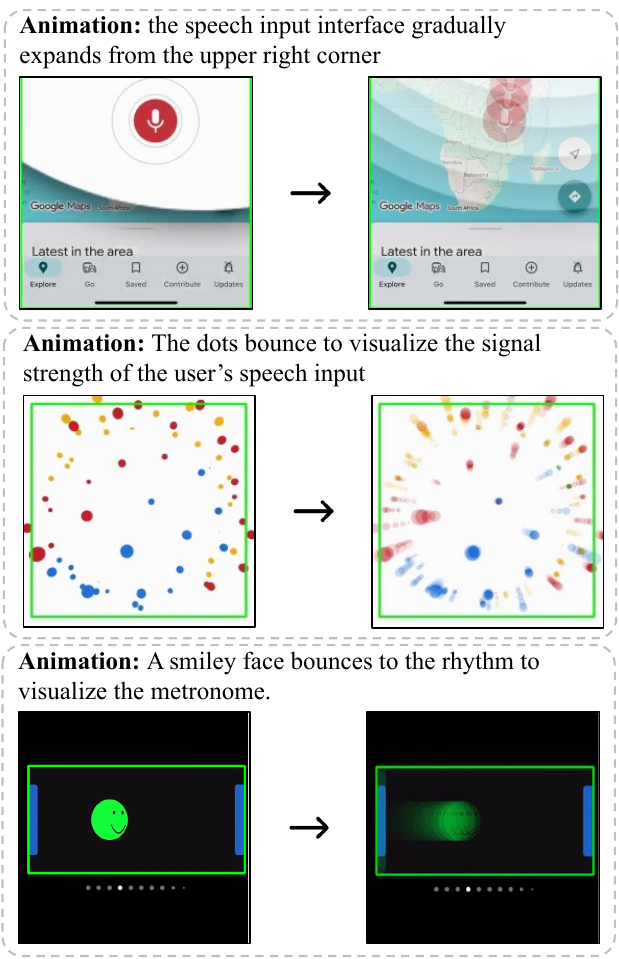}
  \caption{Examples of regular frames vs. motion blended images where motion blended images show the movement patterns in the past few frames.}
  \label{fig:factor_mhi}
\end{figure}

\subsection{Performance on Video Models}
We conducte additional evaluations using Video LLaMa, LLaVA-Video, Qwen-2.5-VL, and Gemini-2.5-pro, where we use videos directly as input. The results are listed in \autoref{tab:video_rq3}.
\begin{table}[t]
\resizebox{\linewidth}{!}{
\begin{tabular}{lccc}
\toprule
Model & Mean & Var & Score distribution (0--5) \\
\midrule
gemini-2.5-pro (I)          & 2.66 & 1.57 & \includegraphics[width=6cm,height=0.3cm]{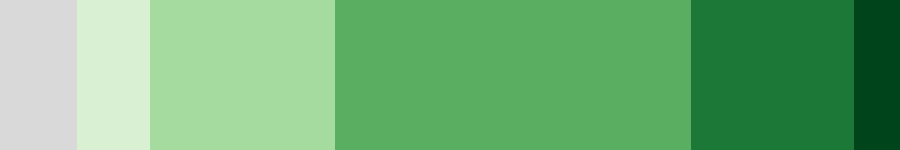} \\
gemini-2.5-pro (S)          & 3.28 & 1.01 & \includegraphics[width=6cm,height=0.3cm]{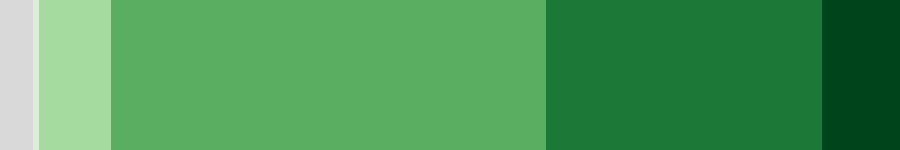} \\
qwen2.5-vl-72b-instruct (I) & 2.44 & 1.53 & \includegraphics[width=6cm,height=0.3cm]{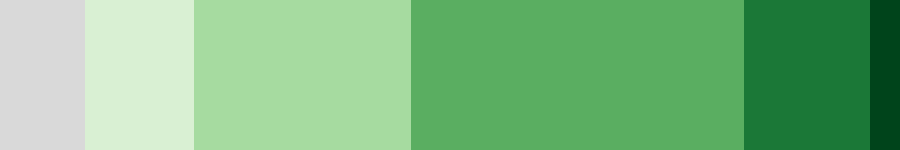} \\
qwen2.5-vl-72b-instruct (S) & 3.02 & 1.55 & \includegraphics[width=6cm,height=0.3cm]{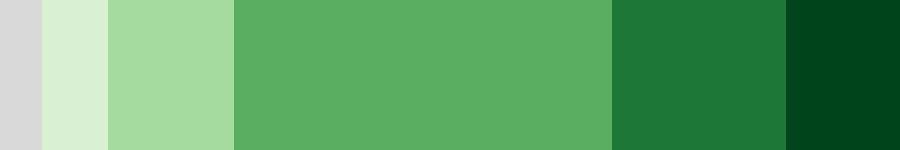} \\
llava-video-7b-local (I)    & 2.06 & 1.39 & \includegraphics[width=6cm,height=0.3cm]{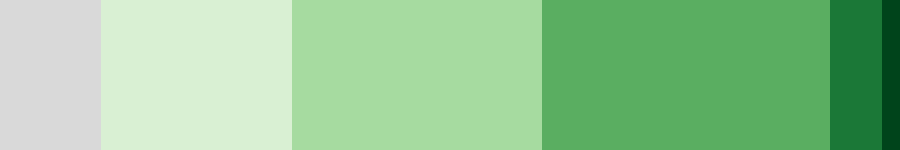} \\
llava-video-7b-local (S)    & 2.29 & 1.45 & \includegraphics[width=6cm,height=0.3cm]{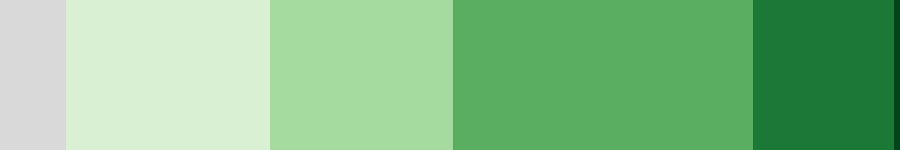} \\
videollama3-7b-local (I)    & 1.50 & 1.26 & \includegraphics[width=6cm,height=0.3cm]{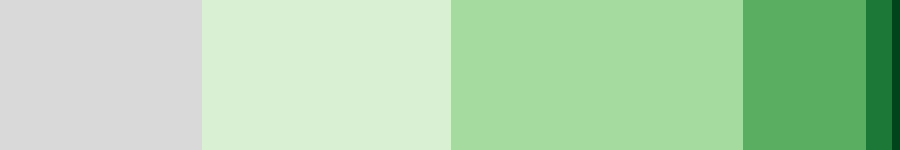} \\
videollama3-7b-local (S)    & 1.47 & 1.32 & \includegraphics[width=6cm,height=0.3cm]{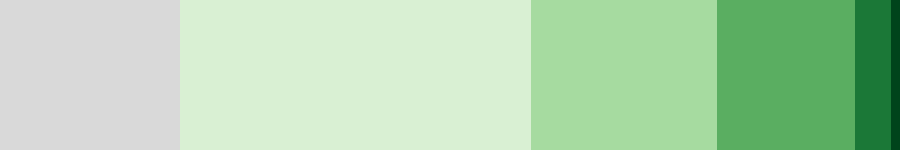} \\
\bottomrule
\end{tabular}}
\caption{Video input performance. (I): individually compared. (S): compared to summary.}
\label{tab:video_rq3}
\end{table}

\section{Additional Details for \textit{MCPC}}
\label{ap:mcp}
\subsection{MCPC Setup Details}

\paragraph{Motion.} To explicitly capture temporal dynamics, we generate a simplified recency-weighted blended motion image inspired by Motion History Image \cite{JimDavis2001MHI}, which integrates changes across multiple frames into a single static representation. This is used as a unified technique to encode motion for models that does and does not have native temporal processing capabilities. The blended image is computed as:
\[
B \;=\; \frac{1-\gamma}{1-\gamma^{N}} \sum_{k=1}^{N} \gamma^{\,N-k}\, F_k
\]
where $B$ is the blended motion image, $F_k\!\in\!\mathbb{R}^{H\times W\times C}$ is the $k$-th frame (indexed $k=1$ oldest $\rightarrow$ $k=N$ newest), $N$ is the number of frames, and $\gamma$ is the exponential decay factor set as 0.85 giving higher weight to recent frames (operations are elementwise over pixels/channels). This representation visualizes temporal changes, such as trajectories, transitions, and rotations. In our implementation, we create blended images at 10 fps, where each blended image blends the 6 most recent frames sampled at 60 fps. Example outcomes are illustrated in \autoref{fig:factor_mhi}.

\paragraph{Context.} We evaluate the impact of contextual information by appending textual context description and the user interaction description to the model's prompt. While this information was included by default in prior evaluations, we explicitly varied this factor here to quantify its impact on performance.

\paragraph{Perceptual Caption.} Perceptual captions are human-annotated textual descriptions of the animation effects, which function as ``alt text'' for the visual dynamics. This setup tests the hypothesis that if a model struggles with raw motion perception, providing an explicit textual description of the movement will bridge the perception gap and improve reasoning performance.

\subsection{Statistical Significance Test}
\label{app-subsec: rq4-stats-test}

Following Appendix \ref{app-subsec: rq2-stats-test} and \ref{app-subsec:rq3-stats-test}, we adopt the McNemar's test and the Wilcoxon signed-rank test for the purpose categorization (RQ2) and UI animation interpretation (RQ3), respectively.
\Cref{tab: rq4_rq2_stats_test,tab: rq4_rq3_stats_test} report the results, respectively.
For purpose categorization (RQ2), the improvement introduced by \textit{MCP} is not statistically significant. 
In contrast, for interpretation (RQ3), \textit{MCP} yields a statistically significant improvement.

\end{document}